\documentclass[10pt,twocolumn,letterpaper]{article}

\usepackage{times}
\usepackage{graphicx}

\usepackage[labelfont=bf]{caption}
\usepackage{multirow}
\usepackage{blindtext}
\usepackage{booktabs}
\usepackage{makecell}
\usepackage{siunitx}

\usepackage{soul}

\usepackage{amsmath,amssymb,amsfonts}
\usepackage{eucal}
\usepackage{enumitem}
\usepackage{algorithm}
\usepackage{algcompatible}
\algnewcommand\algorithmicreturn{\textbf{return}}
\algnewcommand\RETURN{\State \algorithmicreturn}

\usepackage[pagebackref=true,breaklinks=true,colorlinks,bookmarks=false]{hyperref}
\usepackage[square,numbers,sort]{natbib}
\usepackage[capitalize]{cleveref}  

\newcommand{\noiserange}{\mathcal{R}(\sigma_{tr})}
\newcommand{\testnoise}{\sigma_{test}}

\newcommand{\RV}[1]{{#1}}

\def\abstract
   {%
   \centerline{\large\bf Abstract}%
   \vspace*{12pt}%
   \it%
   }

\begin{document}

\title{Noise2Recon: Enabling Joint MRI Reconstruction and Denoising with Semi-Supervised and Self-Supervised Learning}

\author{Arjun D Desai\thanks{Equal contribution}\textsuperscript{$\;\;$,}\thanks{\href{mailto:arjundd@stanford.edu}{\texttt{arjundd@stanford.edu}}} \and Batu M Ozturkler\footnotemark[1] \and Christopher M Sandino \and Robert Boutin \and Marc Willis \and Shreyas Vasanawala \and Brian A Hargreaves \and Christopher R\'e \and John M Pauly \and Akshay S Chaudhari\\
}

\date{Stanford University}

\maketitle
\vspace{-2em}

\begin{abstract}
   Deep learning (DL) has shown promise for faster, high quality accelerated MRI reconstruction. However, supervised DL methods depend on extensive amounts of fully-sampled (labeled) data and are sensitive to out-of-distribution (OOD) shifts, particularly low signal-to-noise ratio (SNR) acquisitions. To alleviate this challenge, we propose Noise2Recon, a model-agnostic, consistency training method for joint MRI reconstruction and denoising that can use both fully-sampled (labeled) and undersampled (unlabeled) scans in semi-supervised and self-supervised settings. With limited or no labeled training data, Noise2Recon outperforms compressed sensing and deep learning baselines, including supervised networks, augmentation-based training, fine-tuned denoisers, and self-supervised methods, and matches performance of supervised models, which were trained with 14x more fully-sampled scans. Noise2Recon also outperforms all baselines, including state-of-the-art fine-tuning and augmentation techniques, among low-SNR scans and when generalizing to other OOD factors, such as changes in acceleration factors and different datasets. Augmentation extent and loss weighting hyperparameters had negligible impact on Noise2Recon compared to supervised methods, which may indicate increased training stability. Our code is available at \url{https://github.com/ad12/meddlr}.
\end{abstract}

\section{Introduction}
MRI is a non-invasive imaging modality with high diagnostic quality owing to its excellent soft-tissue contrast. However, because data acquisition can be inherently slow, MRI suffers from long scan times, and thus, requires accelerated imaging techniques to enable clinical applications. One such approach is parallel imaging (PI), where redundant measurements among receiver coils are used to resolve coherent aliasing artifacts from uniformly undersampled data \cite{pruessmann1999sense, griswold2002generalized}. Another powerful tool to reconstruct undersampled k-space data is compressed sensing (CS), which exploits the sparsity of the reconstructed image in a hand-crafted transform domain \cite{cs-mri}. However, PI methods often have limited efficacy at large acceleration factors, while CS techniques have long reconstruction times due to their iterative nature and require careful fine-tuning of hyperparameters.

Deep-learning (DL) methods have shown potential for enabling higher acceleration factors than PI and CS methods and for improving the quality of the reconstructed images \cite{hammernik2018learning,sandino2020compressed,wang2016accelerating}. The success of these methods can be attributed to their ability to effectively regularize the MRI reconstruction problem and provide much faster reconstruction times compared to CS, which is critical for increasing clinical throughput.

Despite the preliminary success of DL-based methods in MRI reconstruction methods, several challenges remain prior to their widespread clinical adoption of these approaches. One such challenge is their dependence on large amounts of fully-sampled (i.e. \textit{labeled\footnote{Fully-sampled MRI scans provide supervisory signals via labels to compute regression losses.} data}) training data. Given long scan times for fully-sampled scans, MRI acquisitions are routinely accelerated in clinical practice \cite{liu2018highly,cheng2016comprehensive,chaudhari2019combined}. While there are often more accelerated scans than fully-sampled scans, supervised DL reconstruction methods can only utilize fully-sampled scans for training. In such scenarios where fully-sampled images are scarce or absent, techniques that can leverage information from clinically available undersampled datasets are desirable.

\begin{figure}[ht!]
    \centering
  \begin{center}
      \includegraphics[width=\linewidth]{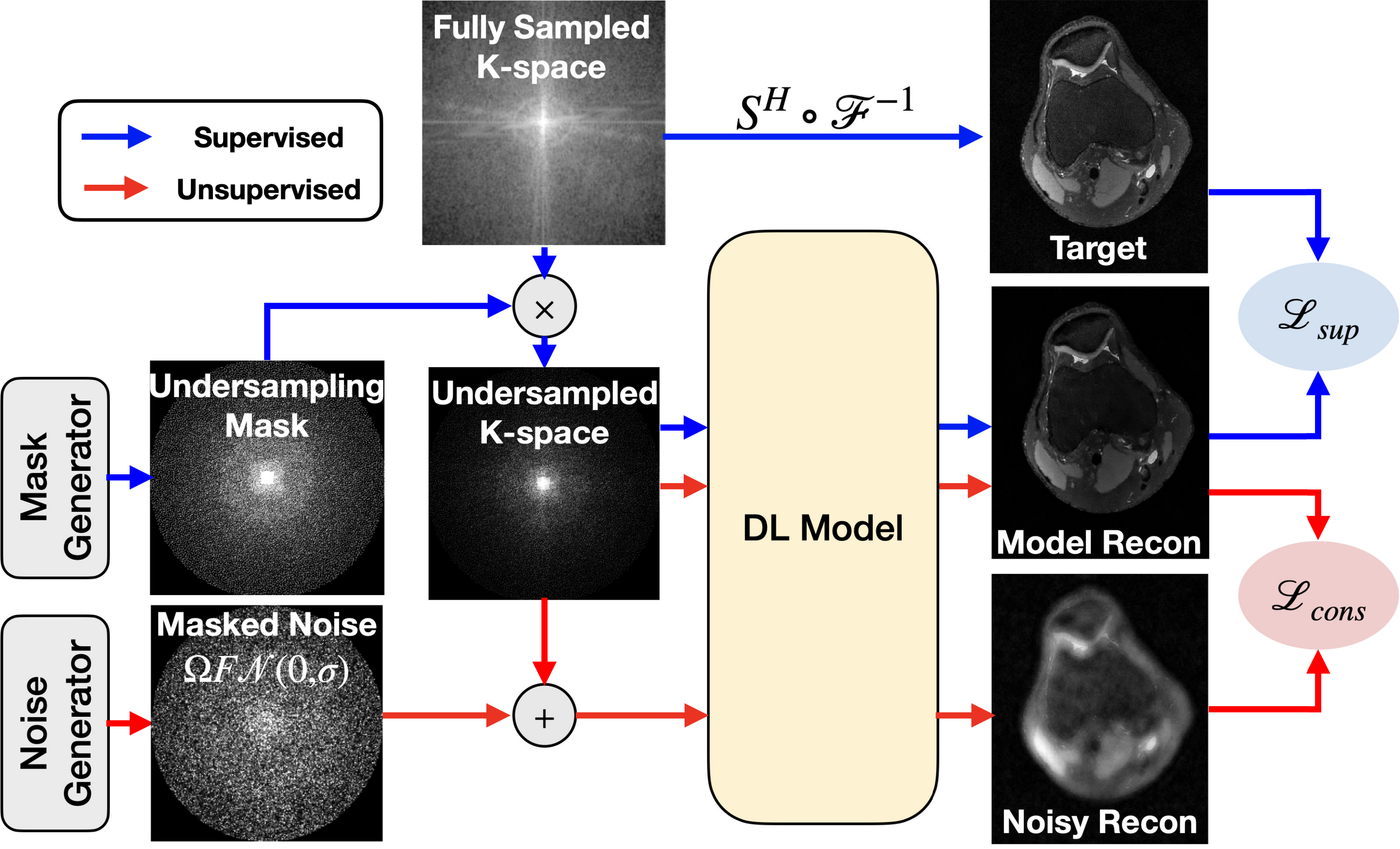}
  \end{center}
  \caption{The Noise2Recon schematic for label-efficient joint reconstruction and denoising. In the semi-supervised setup, fully-sampled scans follow the supervised training paradigm (blue arrows), \RV{where scans are retrospectively undersampled, reconstructed by the model $f_\theta$, and optimized with respect to the available ground-truth reference (i.e. \textit{target})}. Undersampled scans (prospectively undersampled with mask $\Omega$) are augmented with masked, zero-mean complex Gaussian noise with standard deviation $\sigma$, which is sampled from a predefined range. The same model \RV{$f_\theta$} reconstructs both the non-augmented and augmented scans. The reconstruction of the non-augmented scan is used as a pseudo-label for the reconstruction of the augmented scan, a process which we refer to as \textit{consistency}. The total loss is a weighted sum of the supervised and consistency losses: $\mathcal{L}_{total} = \mathcal{L}_{sup} + \lambda\mathcal{L}_{cons}$. Self-supervised Noise2Recon (Noise2Recon-SS) replaces the supervised pathway with self-supervised training from \cite{Yaman_self}.}
  \label{fig:n2r-schematic}
\end{figure}

Additionally, both supervised DL and CS reconstruction techniques, are sensitive to data distribution shifts induced by common perturbations during data acquisition and changes in scan parameters \cite{Chaudhari_2021,hammernik2021systematic}. Previous work has shown that small structural perturbations can result in amplified artifacts among images reconstructed with DL and CS-based methods \cite{darestani21a, Lustig_Donoho_Pauly_2007,Antun30088}. Given the heterogeneity of MR hardware and sequence configurations, one common perturbation that current reconstruction algorithms are vulnerable to is noise, which can vary considerably among different scans. For iterative CS methods, the maximum acceleration factor for reasonable signal recovery is bounded by measurement noise \cite{Donoho_Maleki_Montanari_2011}. Thus, CS-based MRI reconstructions might fail to converge to a feasible solution in high noise regimes \cite{Virtue_Lustig_2017}. The reconstruction quality of DL-based methods also degrades considerably when a deviation in SNR between training and testing is present \cite{Knoll2019b}. Changes in acquisition parameters, such as acceleration factor, can also present challenges for DL reconstruction networks. Recent work has explored robustness to such distribution shifts resulting from anatomical changes \cite{jalal2021robust,raj2020improving}, but these methods do not consider robustness to routine perturbations in the noise of observed signals.

Motivated by these challenges of data paucity and robustness to distribution shifts, we propose Noise2Recon, a label-efficient DL method that performs joint MRI reconstruction and denoising. \RV{Noise2Recon combines regularization properties of consistency training \cite{uda, sohn2020fixmatch} and denoising \cite{batson2019noise2self, moran2020noisier2noise} to provide label-efficient, SNR-robust MRI reconstruction.} In Noise2Recon, available fully-sampled scans are used to train a model with respect to a conventional supervised MRI reconstruction objective. For each undersampled-only scan (no fully-sampled reference), Noise2Recon generates reconstructions for both the undersampled scan and a noise-augmented rendition of the same scan. A consistency loss is used between the clean and noisy reconstructions to enforce the model to be noise-invariant. A schematic of our method is shown in \cref{fig:n2r-schematic}.

At its core, Noise2Recon's consistency framework can utilize \textit{both} fully-sampled and undersampled scans to simultaneously enable reconstruction in label-limited settings and to increase robustness to noise. The advantage of the consistency-training-based formulation is that no assumptions are required on the statistical properties of the input signal to reconstruct in contrast to existing data-efficient denoising approaches \cite{lehtinen2018noise2noise, batson2019noise2self,hendriksen2020noise2inverse}. Furthermore, \RV{Noise2Recon is model-agnostic and can be extended to unsupervised settings, where no fully-sampled references are available}. With these benefits, the main contributions of our work are as follows: 
\vspace{-0.5em}
\begin{enumerate}
    \item We propose Noise2Recon, a model-agnostic, label-efficient framework for joint MRI reconstruction and denoising using consistency-based training via noise augmentations. 
    \item We demonstrate Noise2Recon outperforms state-of-the-art CS and DL supervised and self-supervised baselines in label-limited settings for both feed-forward and unrolled architectures. Among 12x and 16x retrospectively undersampled 3D fast spin echo (FSE) knee scans, Noise2Recon outperformed baselines by up to +0.055 structural similarity (SSIM), +0.84dB peak signal-to-noise ratio (pSNR), and $-$0.032 nRMSE.
    \item We show \RV{that} Noise2Recon increases robustness for reconstructing images in out-of-distribution, noisy acquisitions by up to \RV{+0.08 SSIM}, +0.82dB pSNR, \RV{and $-$0.077 nRMSE} compared to standard augmentation and fine-tuning approaches.
    \RV{\item We build a self-supervised variant of Noise2Recon (termed \textit{Noise2Recon-SS}) that can be trained without any fully-sampled references (i.e. unsupervised settings). Noise2Recon-SS is competitive with self-supervised baselines among in-distribution, high-SNR data and outperforms these methods among noisy acquisitions.}
\end{enumerate}
All code, experimental configurations, and pretrained models are openly available\footnote{\url{https://github.com/ad12/meddlr}}.

\section{Related Work}
In this section, we outline existing supervised image reconstruction, data-limited image reconstruction and image denoising methods that motivated our method.

\subsection{Supervised Image Reconstruction}
The vast majority of image recovery methods perform learning in a supervised fashion, where a large dataset consisting of fully-sampled (labeled) examples is needed to perform training. Supervised DL approaches either directly invert the forward imaging model with a feed-forward convolutional neural network (CNN) \cite{zhang2017, 2017deepcnn, akccakaya2018scan, quan2018compressed} or unroll an iterative algorithm, which alternates between a data-fidelity step and a CNN-based regularization step \cite{pezzotti2020adaptive,adler2018learned,Meinhardt_Moeller_Hazirbas_Cremers_2017,Aggarwal_Mani_Jacob_2019, hammernik2018learning, Yaman_self}. However, these methods require large number of fully-sampled scans and are not designed to leverage undersampled scans.

\subsection{Data-Efficient Image Reconstruction}
The supervised data-dependence problem is not unique to MRI reconstruction -- in fact, several data-efficient methods for general image recovery have been proposed in prior computer vision techniques \cite{lehtinen2018noise2noise, hu2021system, batson2019noise2self, hendriksen2020noise2inverse,romano2017little}. Recently, these approaches have motivated data-efficient methods for MRI reconstruction. \citet{Liu_Sun_Eldeniz_Gan_An_Kamilov_2020} extend regularization by denoising (RED) \cite{romano2017little} to utilize priors from more general artifact removal networks to train with prospectively undersampled data. \cite{darestani2021accelerated} use untrained networks to incorporate the architecture of a CNN as an image prior. Generative adversarial networks using unpaired datasets \cite{Lei_Mardani_Pauly_Vasanawala_2021} or only undersampled datasets \cite{cole2020unsupervised} have also shown promise for data-efficient MRI reconstruction. Other methods involve self-supervised learning \cite{Yaman_self, yaman2021zero, demirel202120} and dictionary-based learning \cite{Lahiri_Wang_Ravishankar_Fessler_2021,ravishankar2017efficient,song2019coupled} to enable reconstruction when fully-sampled scans are limited. Recently, image-based augmentations were also verified to help decrease data dependence for fully-supervised networks \cite{fabian2021data}. While these data-efficient approaches reduce dependence on fully-sampled data, they, like supervised DL and CS reconstruction methods, are sensitive to data distribution shifts induced by common perturbations during data acquisition and changes in scan parameters.

\subsection{Data-Efficient Image Denoising}
Recently, there have been several approaches proposed for image denoising problems that do not require access to a large dataset with fully-sampled references. Before DL approaches, plug-and-play priors \cite{pnp} and iterative image denoising priors \cite{romano2017little} were shown to be very effective in a wide range of inverse problems. \cite{lehtinen2018noise2noise} showed that image recovery with neural networks can be performed without ground-truth images by only using images corrupted by zero-mean noise. Reference-less denoising methods have also extended to self-supervised training using only noisy images to model denoising \cite{batson2019noise2self,hu2021system} and other imaging inverse problems \cite{hendriksen2020noise2inverse}. However, these methods operate under the assumption that noise exhibits statistical independence across different dimensions of the measurements.

\section{Preliminaries}
In this section, we first introduce the operating notation for the reconstruction problem (summarized in \cref{tab:glossary}). We then formalize the optimization for supervised MRI reconstruction and for unsupervised denoising. Finally, we introduce our proposed label-efficient method for joint MRI reconstruction and denoising.

\subsection{Notation}
We consider the multi-coil accelerated MRI acquisition setup, where the observed k-space samples are acquired across multiple receiver coils. The forward model for this problem can be formulated as follows:
\begin{equation}
    \RV{y = \Omega FSx^* + \tilde{\epsilon}}
\end{equation}
where $y$ is the set of observed, complex-valued measurements in k-space for all coils, $x^*$ is the true image we would like to reconstruct, $S$ is the set of sensitivity maps associated with each receiver coil, $F$ is the Fourier transform matrix, and $\Omega$ is the k-space undersampling mask. \RV{$\tilde{\epsilon}$} is the masked additive complex Gaussian noise resulting from thermal noise \cite{macovski1996noise}. $\tilde{\epsilon}$ is the same dimension as $y$.

Consider a dataset $\mathcal{D}$ that consists of scans with fully-sampled (supervised) k-space data ($\mathcal{D}^{(s)}$) and scans with undersampled-only (unsupervised) k-space data ($\mathcal{D}^{(u)}$) --- i.e. $\mathcal{D} = \mathcal{D}^{(s)} \cup \mathcal{D}^{(u)}$. $y^{(s)}_i \in \mathcal{D}^{(s)}$ and $y^{(u)}_j \in \mathcal{D}^{(u)}$ are the k-space measurements of the $i^{th}$ example in the supervised dataset and $j^{th}$ example in the unsupervised dataset, respectively. $x^{(s)}_i$ is the image space counterpart of $y^{(s)}_i$, and $x^{(j)}_i$ is the image space counterpart of $y^{(j)}_i$. $f_{\theta}$ is the model parameterized by $\theta$ trained to reconstruct images from undersampled k-space data. The operator $|\cdot|$ denotes the cardinality (i.e. size) of the dataset. In most practical clinical scenarios where accelerated imaging is used, $|\mathcal{D}^{(s)}| << |\mathcal{D}^{(u)}|$.

\begin{algorithm}[t!]
 \caption{Noise2Recon's main learning algorithm.}
 \label{alg:n2r-method}
 \begin{algorithmic}[1]
 \REQUIRE dataset $\mathcal{D}=\mathcal{D}^{(s)}\cup\mathcal{D}^{(u)}$, model $f_\theta$
 \REQUIRE batch size N, constant $\sigma$, constant $\lambda$
 
 \FOR {\text{sampled minibatch} $\{(y_i^{(s)}, x_i^{(s)})\}_{i=1}^{N_s}$, $\{(\Omega_{y_j^{(u)}}, y_j^{(u)})\}_{j=1}^{N - N_s}$}
    \STATE $N_u \gets N - N_s$ \Comment{Num. unsupervised examples in batch}
    \FORALL {$i \in \{1, \cdots, N_s\}$}
        \STATE $\hat{x}^{(s)}_i \gets f_\theta(y_i^{(s)})$
    \ENDFOR
    \FORALL {$j \in \{1, \cdots, N_u\}$}
        \STATE $\hat{x}^{(u)}_j \gets f_\theta(y_j^{(u)})$
        \STATE $\epsilon_j \in \mathbb{C}^{shape(y_j^{(u)})} \sim \mathcal{N}(0, \sigma)$
        \STATE $\tilde{\epsilon_j} \gets \Omega_{y_j^{(u)}} \epsilon_j$
        \STATE $\tilde{x}^{(u)}_j \gets f_\theta(y_j^{(u)} + \tilde{\epsilon_j})$
    \ENDFOR
    \STATE $\mathcal{L}_{total} \gets \frac{1}{N_s} \sum_{i=1}^{N_s}\mathcal{L}_{sup}(\hat{x}^{(s)}_i, x^{(s)}_i) + \frac{\lambda}{N_u}\sum_{j=1}^{N_u}\mathcal{L}_{cons}(\tilde{x}^{(u)}_j, \hat{x}^{(u)}_j)$
    \STATE update network $f_\theta$ to minimize $\mathcal{L}_{total}$
 \ENDFOR
 \RETURN{} $f_\theta$
\end{algorithmic} 
\end{algorithm}

\subsection{Supervised MRI Reconstruction}
\label{sec:supervised-methods}
In supervised MRI reconstruction, training is performed using only data where fully-sampled references exist (i.e. $\mathcal{D}^{(s)}$). In these cases, an undersampled input can be simulated by sampling an undersampling mask $\Omega$ from a distribution of undersampling patterns and applying {this mask} to the fully-sampled k-space $y^{(s)}_i$. As fully-sampled scans can be retrospectively undersampled, different masks can be generated for different inputs. End-to-end training of model $f_{\theta}$ minimizes
\begin{equation}
    \min_{\theta} \frac{1}{|\mathcal{D}^{(s)}|} \sum_{\RV{i=0}}^{|\mathcal{D}^{(s)}|} \mathcal{L}_{sup}(f_{\theta}(\RV{\Omega y^{(s)}_i, A_i^H}), x^{(s)}_i)
\end{equation}
where $\mathcal{L}_{sup}$ is a supervised loss function and \RV{$A_i^H$ is the Hermitian of the imaging model (includes mapping from k-space to image space) for the $i^{th}$ example. $f_\theta$ can be any learnable parameterized model, such as feed-forward or unrolled networks.}

To avoid overfitting in data-scarce settings, supervised reconstruction methods can use data augmentation to simulate larger labeled training datasets \cite{fabian2021data}. \RV{These augmentations can either be performed in image space (e.g. rotation, scaling, shifting, etc.) or in k-space (e.g. additive noise). For simplicity, we consider a single augmentation $T$ applied to k-space with probability $p$.} Augmentations performed in k-space are often label-invariant -- \RV{i.e. the ground-truth reference image (label) should not change as a result of applying the augmentation (\cref{fig:supervised-aug-schematic})}. The loss for example $x^{(s)}_i$ with k-space augmentations can be written as
\begin{equation}
    \mathcal{L}_{sup}(f_{\theta}(\Omega T_p(y^{(s)}_i), A_i^H), x^{(s)}_i) 
\end{equation}

\subsection{Unsupervised Image Denoising}
Unsupervised denoising techniques can be formulated by selecting an example \RV{$y^{(u)}_i$} from an unsupervised dataset \RV{$|\mathcal{D}^{(u)}|$}, corrupting the example with a known or expected signal corruption model $\Psi$, and training a model to recover the original signal \RV{$y^{(u)}_i$} from the corrupted signal \RV{$\Psi(y^{(u)}_i)$}. More formally this can be written as

\RV{
\begin{equation}
    \min_{\theta} \frac{1}{|\mathcal{D}^{(u)}|} \sum_{i=0}^{|\mathcal{D}^{(u)}|} \mathcal{L}_{unsup}(f_{\theta}(\Psi(y^{(u)}_i), y^{(u)}_i)
\end{equation}
}

where $\mathcal{L}_{\RV{unsup}}$ is an arbitrary regression loss function (e.g. $\ell_1, \ell_2$), and $\Psi$ denotes the corruption model with noise drawn from a predefined distribution (typically zero-mean Gaussian) $\mathcal{N}$ (\RV{first step in \cite{moran2020noisier2noise}}). The unsupervised loss function can be modified to incorporate pairs of corrupted images from the same example with independently sampled noise \cite{lehtinen2018noise2noise}, or only the corrupted image alongside an explicit corruption model \cite{batson2019noise2self}. \RV{By observing the posterior distribution of clean images under corrupted images, these techniques can be extended to images that are corrupted by any exponential-family distribution \cite{kim2021noise2score}.}

\subsection{Proposed Method: Noise2Recon}
\label{sec:method-n2r}
Current supervised reconstruction methods achieve state-of-the-art results with large amounts of fully-sampled data, but these methods are prone to overfitting in data-scarce settings. Model regularization techniques, such as $\ell_1$/$\ell_2$ regularization and dropout \cite{srivastava2014dropout}, can help mitigate overfitting. However, these methods are based predominantly on prior-driven assumptions about model weights (e.g. sparsity). Given that fully-sampled data can often be scarce, reconstruction methods that can leverage a mixture of fully-sampled and prospectively undersampled data and \RV{can incorporate} data-driven regularization would be helpful. Additionally, while both denoising and reconstruction tasks are critical for recovering high quality images, they are formulated as disjoint, sequential operations. The separation of these objectives may be optimal for each task individually, but may lead to poor optimization for both tasks jointly.

In this work, we propose a label-efficient method for joint MR reconstruction and denoising that mitigates overfitting in data-scarce settings and increases robustness to noisy OOD acquisitions. In the semi-supervised setting, Noise2Recon complements the supervised training paradigm described in \cref{sec:supervised-methods} by adding a noise-augmentation consistency training paradigm (Fig.\ref{fig:n2r-schematic}). Examples without fully-sampled references (unsupervised) are augmented with masked additive noise. The model $f_{\theta}$ generates reconstructions for both unsupervised images ($f_{\theta}(y^{(u)})$) and noise-augmented unsupervised images ($f_{\theta}(y^{(u)}+\Omega_{y^{(u)}}\epsilon)$), where $\Omega_{y^{(u)}}$ is the undersampling mask that was used to acquire unsupervised example $y^{(u)}$. A consistency loss ($\mathcal{L}_{cons}$) is enforced between reconstructions of the unsupervised examples and their noise-augmented counterparts to build noise-invariant reconstruction models. End-to-end training with Noise2Recon seeks to mimimize a weighted sum of the supervised loss ($\mathcal{L}_{sup}$) and the unsupervised consistency loss ($\mathcal{L}_{cons}$). Thus, the objective can be written as
\begin{align}
\label{eq:n2r-obj}
\begin{split}
	\min_\theta \;&\mathbb{E}[\mathcal{L}_{sup}(f_{\theta}(\Omega y^{(s)}), x^{(s)})] \\+ \lambda &\mathbb{E}[\mathcal{L}_{cons}(f_{\theta}(y^{(u)}+\epsilon),f_{\theta}(y^{(u)}))]
\end{split}
\end{align}
where undersampling mask $\Omega$ can be randomly generated for fully-sampled data, $\lambda$ is a weighting constant, and $\epsilon$ is a randomly generated noise map drawn from a complex-Gaussian distribution with standard deviation $\sigma$. \cref{alg:n2r-method} summarizes the proposed method.

\paragraph{Simulating noise for consistency augmentations} Noise in MRI is dominated by thermal fluctuations in the subject and the receiver electronics \cite{redpath1998signal}. This noise source can be modeled as additive complex-valued Gaussian noise added to each acquired k-space sample \cite{macovski1996noise}. Thus, for unsupervised example ${y^{(u)}_j}$, we generate \RV{masked complex-gaussian noise $\tilde{\epsilon}_j = \Omega_{y^{(u)}_j} \epsilon_j$}, where noise map $\epsilon_j \sim \mathcal{N}(0,\sigma_{tr})$ and $\mathcal{N}$ is a zero-mean complex-gaussian distribution with standard deviation $\sigma_{tr}$. $\sigma_{tr}$ is chosen from a specified range (for training) $\noiserange = [\sigma_{tr}^L, \sigma_{tr}^U)$. \RV{The masked noise map is added to the pre-normalized image so that it induces the same relative change in SNR across scans.} We consider a \RV{pre-whitened coil} setting where noise for separate coils is independent and identically distributed. In the case where correlation between noise for independent coils is present, noise pre-whitening can be performed as a preprocessing step to ensure that in our framework the encountered noise distribution is uncorrelated \cite{hansen2015image}.


\paragraph{Balanced data sampling} As the supervised and consistency objectives are computed over a disjoint set of examples, the weighting of each objective \RV{across the full dataset $\mathcal{D}$} is governed by the rate of sampling from $\mathcal{D}^{(s)}$ and $\mathcal{D}^{(u)}$, respectively. More formally, 

\RV{
\begin{equation*}
 \frac{\nabla_\theta \sum_{i=0}^{|D_s|}\mathcal{L}_{sup}(f_{\theta}(\Omega y_i^{(s)}), x_i^{(s)})}{\nabla_\theta\sum_{j=0}^{|D_u|}\mathcal{L}_{cons}(f_{\theta}(y_j^{(u)}+\epsilon),f_{\theta}(y_j^{(u)}))} \propto \frac{|\mathcal{D}^{(s)}|}{|\mathcal{D}^{(u)}|}.
\end{equation*}
}

In this setting, the optimization is sensitive to the ratio of supervised to unsupervised examples. One solution to this would involve modifying the loss weighting $\lambda$ to account for different relative dataset sizes. However, this solution would require extensive tuning for $\lambda$ and would still perpetuate uneven optimization at different stages of the training cycle.

We propose a balanced data sampling scheme that samples unsupervised and supervised examples at a rate determined by a fixed ratio $T_S$:$T_U$. For every $T_S$ supervised examples that are sampled during training, $T_U$ unsupervised examples are sampled. In this formulation, this sampling method implicitly eliminates the influence of the relative sizes of the supervised and unsupervised datasets on the relative weighting between the supervised and consistency objectives. \cref{alg:balanced-sampling} provides an overview of balanced sampling.

\paragraph{Self-Supervised Noise2Recon (\textit{Noise2Recon-SS})} Our method can also be trivially extended to a fully unsupervised setting, \textit{where fully-sampled scans are not available}. In this setup, the supervised training pathway in Noise2Recon can be replaced with the self-supervised training setup from \cite{Yaman_self}.

\begin{algorithm}[t!]
 \caption{Balanced sampling algorithm for creating a batch.}
 \label{alg:balanced-sampling}
 \begin{algorithmic}[1]
 \REQUIRE supervised dataset $\mathcal{D}^{(s)}$, unsupervised dataset $\mathcal{D}^{(u)}$, model $f_\theta$
 \REQUIRE batch size N, supervised period $T_s$, unsupervised period $T_u$
 \STATE $N_s = \frac{N * T_s}{T_s + T_u}$, $N_u = \frac{N * T_u}{T_s + T_u}$
\FORALL {$n \in \{1, \cdots, N_s\}$}
        \STATE Sample $k \in \{1,...,|D_s|\}$
        \STATE $I_s(n) = k$
\ENDFOR
\FORALL {$m \in \{1, \cdots, N_u\}$}
        \STATE Sample $k \in \{1,...,|D_u|\}$
        \STATE $I_u(m) = k$
\ENDFOR
 \RETURN $\{(y_i^{(s)}, x_i^{(s)}), i \in I_u\}$, $\{(\Omega_{y_j^{(u)}}, y_j^{(u)}), j \in I_s\}$
\end{algorithmic} 
\end{algorithm}

\section{Experiments}
Our goal is to demonstrate whether Noise2Recon can leverage noise augmentations for task-based regularization that can improve performance in both high-SNR and low-SNR settings. We evaluate whether Noise2Recon can (1) outperform supervised and state-of-the-art self-supervised methods in label-scarce scenarios and (2) improve robustness to reconstruction in noisy settings. We conduct extensive ablations to assess the advantages of the consistency objective and the balanced sampling.

\subsection{Dataset}
\label{sec:dataset}
We performed experiments on the publicly available fully-sampled 3D fast-spin echo (FSE) multi-coil knee scans (acquistion matrix $k_x \times k_y \times k_z$=320$\times$320$\times$256) from \url{http://mridata.org} \cite{ong2018mridata}. The dataset of 19 subjects was partitioned into 14 subjects (4480 slices) for training, 2 subjects (640 slices) for validation, and 3 subjects for testing (960 slices). 3D scans were demodulated and decoded using the 1D orthogonal inverse Fourier transform along the readout direction, resulting in a hybrid k-space of dimensions $x \times k_y \times k_z$. Sensitivity maps for each volume were estimated using JSENSE (implemented in SigPy \cite{sigpy}) with a kernel-width of 8 and a 20$\times$20 center k-space auto-calibration region \cite{ying2007joint}. Fully-sampled data were retrospectively undersampled with a 2D Poisson Disc undersampling pattern with the same auto-calibration region. For testing, a unique, deterministic undersampling trajectory was generated for each testing volume using a fixed random seed for reproducibility.

\subsection{Experimental Settings}
\label{sec:methods-simulations}
\paragraph{Label scarcity} To evaluate the performance of different methods in label-scarce settings, scans in the training dataset $\mathcal{D}$ were subsampled. Fully-sampled references were retained for $k$ scans in the training dataset and dropped for the remaining $|\mathcal{D}|-k$ scans. More formally, $\mathcal{D}_k \subset \mathcal{D}$ is the set of $k$ training scans for which fully-sampled references are available. A fixed undersampling mask was generated for each scan not in $\mathcal{D}_k$ (i.e. $ x \in \mathcal{D} \setminus \mathcal{D}_k$) to simulate undersampled, reference-less scans. The extent of label scarcity was simulated with different values of $k$ such that larger subsets are supersets of smaller subsets --- i.e. $\mathcal{D}_1 \subset \mathcal{D}_2 \dots \subset \mathcal{D}_N$.

\paragraph{Noisy data} To characterize how different methods generalize to reconstructing noisy OOD scans, noisy acquisitions were simulated for testing scans . For a given noise level $\testnoise$, an uncorrelated multi-channel masked zero-mean complex-Gaussian noise map was generated and added to the undersampled measurements from each coil. The coil measurements were first scaled by the 95\textsuperscript{th} percentile of the magnitude image such that the addition of the noise map would result in an equal reduction of SNR among all scans. Noise level $\testnoise$ was varied from 0 to 1.0, in 0.1 increments. Sample zero-filled SENSE-reconstructed images at different noise levels are shown in \cref{fig:zf-noise}.

\paragraph{Multiple accelerations} To compare how DL methods generalized to acceleration factors not observed during training (i.e. \textit{unseen} accelerations), DL baselines and Noise2Recon were evaluated on scans that were retrospectively undersampled at multiple different accelerations. Each model was trained with scans undersampled at a fixed acceleration $R_{train}$ and evaluated on testing scans undersampled at accelerations $R_{test}$=8,12,16,20,24.

\RV{
\paragraph{Cross-dataset generalizability}
In practice, distribution shifts originate from multiple sources, such as changes in the sampling pattern, contrast, sequence type, etc. To evaluate how DL methods generalize in cases of other sources of distribution shifts, we evaluate all models, which are trained on the 12x-accelerated mridata 3D FSE knee dataset, on the 2D fastMRI brain dataset \cite{zbontar2018fastmri}. This cross-dataset evaluation considers the scenario of several sources of distribution shift, such as anatomy (knee $\rightarrow$ brain), field strength (3T $\rightarrow$ 1.5T), acceleration factor (12x $\rightarrow$ 4x), and sequence type (3D FSE $\rightarrow$ 2D FSE), among others. \cref{app:experimental-setup} provides details on the dataset and the different sources of distribution shifts.

\paragraph{Learning without labels} For certain scan protocols, acquiring fully-sampled scans is infeasible. We evaluated how our self-supervised model variant, termed Noise2Recon-SS, and the state-of-the-art self-supervised method SSDU performed when no fully-sampled training data was available (i.e. unsupervised, $\mathcal{D}_{k=0} = \emptyset$).
}

\subsection{Baseline Methods}
\label{sec:methods-baseline}

\paragraph{Supervised training} Supervised models were trained both without and with noise augmentations (termed \textit{Supervised}, \textit{Supervised+Aug}). All augmentations were performed online (i.e. dynamically during training). Augmentations were designed to be equivalent to those used in comparable Noise2Recon configurations and were applied with a probability of $p$=0.2 (see \cref{app:baseaug-noise-prob} for hyperparameter details). In label-scarce settings, all models were trained with only the available fully-sampled scans in the training dataset $\mathcal{D}_k$.

\paragraph{Fine-tuning (\textit{FT}) from denoisers} \RV{Prior work has demonstrated that denoisers are useful regularizers for general families of inverse problems \cite{romano2017little, lehtinen2018noise2noise, mataev_2019_deepred, reehorst_2019_regularization}. Thus, fine-tuning from pretrained denoising networks may reduce the learning requirements for the reconstruction task while preserving denoising properties of the network, which are critical for generalizing to low-SNR settings. In this baseline, a self-supervised denoising training protocol proposed in \cite{batson2019noise2self} was used to train a denoising model. The resulting model was  fine-tuned on the reconstruction task in a supervised manner, without (\textit{Supervised (FT)}) and with (\textit{Supervised+Aug (FT)}) noise augmentations. Training and configuration details are provided in \cref{app:pretrained-denoisers-implementation}.}

\paragraph{Self-supervision with Data Undersampling (SSDU)} We compared Noise2Recon to a state-of-the-art self-supervision with data undersampling (SSDU) reconstruction baseline \cite{Yaman_self}. While SSDU was designed for training with only undersampled scans, we proposed an extension to adapt it to the semi-supervised setting to ensure fair comparison to Noise2Recon, which is a semi-supervised method. \RV{We also find that SSDU is sensitive to data consistency, which is absent in feed-forward networks (e.g. U-Net). Thus, we include a hard-data consistency post-processing step when using SSDU with feed-forward networks.} Details of this extension, training configuration, and postprocessing are provided in \cref{app:ssdu}.

\paragraph{Compressed sensing (CS)} We included compressed sensing with $\ell_1$-wavelet regularization \cite{Lustig_Donoho_Pauly_2007}, a clinically used scan-specific, iterative reconstruction method, as an additional baseline. Reconstruction was performed slice-by-slice using SigPy where the proximal gradient method was run for \RV{100 iterations} \cite{sigpy}. Details on selection of the regularization parameter $\lambda$ are provided in \cref{app:compressed-sensing-implementation}.

\subsection{Implementation Details}
\label{sec:methods-implementation}
All DL approaches were trained end-to-end using the U-Net architecture implemented in the fastMRI challenge \cite{muckley2020state, ronneberger2015u}. \RV{To characterize whether different methods were model dependent, supervised, SSDU, and Noise2Recon methods were also trained using a proximal gradient descent (PGD) unrolled architecture. For unsupervised experiments, models were also trained with unrolled architecture. \cref{app:hyperparameters} provides architecture and hyperparameter details.}

Models were trained on zero-filled, SENSE-reconstructed complex images generated using the estimated sensitivity maps described in \S\ref{sec:dataset}. Complex images were represented with two-channels corresponding to the real and imaginary components. Inputs were normalized by the 95\textsuperscript{th}-percentile of the image magnitude. To preserve the magnitude distribution during metric computation at inference, outputs of the model were scaled by the normalizing constant. \RV{All experiments were performed with the PyTorch library \cite{paszke2019pytorch}}.
\begin{table}[t]
\centering
\begin{center}
\caption{Mean (std. dev.) performance at different accelerations ($R$) of different reconstruction methods trained with 1 fully-sampled scan ($k=1$) and 13 undersampled scans \RV{using the feed-forward U-Net architecture}. Best performing method at each acceleration is \textbf{bolded}.}
\label{tbl:label-scarcity-k1}
\resizebox{\columnwidth}{!}{
\begin{tabular}{llccc}
\toprule
\textbf{$R$} & \textbf{Method} & \textbf{nRMSE ($\downarrow$)} & \textbf{SSIM ($\uparrow$)} & \textbf{pSNR (dB) ($\uparrow$)}\\
\midrule
\multirow{7}{*}{12x} & Compressed Sensing \cite{Lustig_Donoho_Pauly_2007,sigpy} &  0.175 (0.012) &  0.846 (0.012) &  37.3 (0.3) \\
 & Supervised &  0.162 (0.007) &  0.827 (0.031) &  38.0 (0.2) \\
 & Supervised (FT) &  0.157 (0.015) &  0.810 (0.036) &  38.2 (0.6) \\
& Supervised + Aug &  0.163 (0.008) &  0.816 (0.035) &  37.9 (0.3) \\
 & Supervised + Aug (FT) &  0.157 (0.015) &  0.810 (0.037) &  38.2 (0.7) \\
& SSDU \cite{Yaman_self} &  0.162 (0.007) &  0.846 (0.036) &  37.8 (0.5) \\
  & \textbf{Noise2Recon (Ours)} & \textbf{ 0.142 (0.013)} &  \textbf{0.901 (0.018)} &  \textbf{39.1 (0.6)} \\
\midrule
\multirow{7}{*}{16x} & Compressed Sensing \cite{Lustig_Donoho_Pauly_2007,sigpy} &  0.178 (0.013) &  0.847 (0.011) &  37.1 (0.3) \\
  & Supervised &  0.171 (0.009) &  0.810 (0.032) &  37.5 (0.2) \\
  & Supervised (FT) &  0.160 (0.014) &  0.809 (0.037) &  38.0 (0.6) \\
  & Supervised + Aug &  0.172 (0.009) &  0.812 (0.042) &  37.4 (0.3) \\
 & Supervised + Aug (FT) &  0.167 (0.012) &  0.787 (0.039) &  37.7 (0.5) \\
 & SSDU \cite{Yaman_self} &  0.181 (0.016) &  0.844 (0.042) &  37.0 (0.6) \\
  & \textbf{Noise2Recon (Ours)} &  \textbf{0.151 (0.012)} &  \textbf{0.887 (0.018)} &  \textbf{38.6 (0.5)} \\
\hline  
\end{tabular}
}
nRMSE: normalized root-mean-square error, SSIM: structural similarity, pSNR: peak signal-to-noise-ratio.
\end{center}
\end{table}

\subsection{Evaluation}
We report results on three common image quality metrics computed on magnitude images: normalized root-mean-square error (nRMSE), structural similarity (SSIM, range: [0, 1]) \cite{wang2004image}, and peak signal-to-noise ratio (PSNR, dB).

Additional qualitative evaluation on the 3D mridata FSE knee dataset was performed by two board-certified  radiologists (27 years \& 15 years certification). Readers compared the proposed Noise2Recon method with ground-truth fully sampled scans, SSDU, and the supervised DL reconstructions in high-SNR ($\sigma_{test}=0$) and low-SNR ($\sigma_{test}=0.2$) settings. Noise2Recon and SSDU methods were trained with 1 supervised scan, and supervised method was trained with 14 supervised scans. All DL models used the PGD-unrolled network architecture. Readers were blinded to the reconstruction method, and the order of the reconstructions was randomized. All images were scored for aliasing, SNR, and blurring artifacts on a 5-point ordinal scale: 1-- non-diagnostic, 2-- poor, 3-- minimum diagnostic quality, 4-- good, 5-- excellent.

\section{Results}
\label{sec:results}
\begin{table}[t]
\centering
\begin{center}
\caption{\RV{Mean (std. dev.) performance at different accelerations ($R$) of different reconstruction methods trained with 1 fully-sampled scan ($k=1$) and 13 undersampled scans using the proximal gradient descent unrolled architecture. Best performing method at each acceleration is \textbf{bolded}.}}
\label{tbl:label-scarcity-k1-unrolled}
\resizebox{\columnwidth}{!}{
\begin{tabular}{llccc}
\toprule
\textbf{$R$} & \textbf{Method} & \textbf{nRMSE ($\downarrow$)} & \textbf{SSIM ($\uparrow$)} & \textbf{pSNR (dB) ($\uparrow$)}\\
\midrule
\multirow{6}{*}{12x} & Compressed Sensing \cite{Lustig_Donoho_Pauly_2007,sigpy} &  0.175 (0.012) &  0.846 (0.012) &  37.3 (0.327) \\
    & Supervised &  0.129 (0.009) &  0.887 (0.005) &  39.9 (0.436) \\
    & Supervised + Aug &  0.131 (0.009) &  0.905 (0.005) &  39.8 (0.458) \\
    & MRAugment &  0.132 (0.010) &  0.901 (0.004) &  39.7 (0.492) \\
    & SSDU \cite{Yaman_self} &  0.145 (0.012) &  0.905 (0.012) &  38.9 (0.551) \\
    & \textbf{Noise2Recon (Ours)} &  \textbf{0.127 (0.009)} &  \textbf{0.921 (0.003)} &  \textbf{40.0 (0.408)} \\
\midrule
\multirow{6}{*}{16x} & Compressed Sensing \cite{Lustig_Donoho_Pauly_2007,sigpy} &  0.178 (0.013) &  0.847 (0.011) &  37.1 (0.345) \\
    & Supervised &  0.137 (0.010) &  0.895 (0.001) &  39.4 (0.442) \\
    & Supervised + Aug &  0.137 (0.011) &  0.899 (0.003) &  39.4 (0.475) \\
    & MRAugment &  0.141 (0.010) &  0.882 (0.003) &  39.2 (0.438) \\
    & SSDU \cite{Yaman_self} &  0.150 (0.013) &  0.896 (0.008) &  38.6 (0.541) \\
    & \textbf{Noise2Recon (Ours)} &  \textbf{0.135 (0.009)} &  \textbf{0.903 (0.002)} &  \textbf{39.5 (0.386)} \\
\bottomrule
\end{tabular}
}
nRMSE: normalized root-mean-square error, SSIM: structural similarity, pSNR: peak signal-to-noise-ratio.
\end{center}
\end{table}

\begin{figure}[t!]
  \centering
  \begin{center}
      \includegraphics[width=\linewidth]{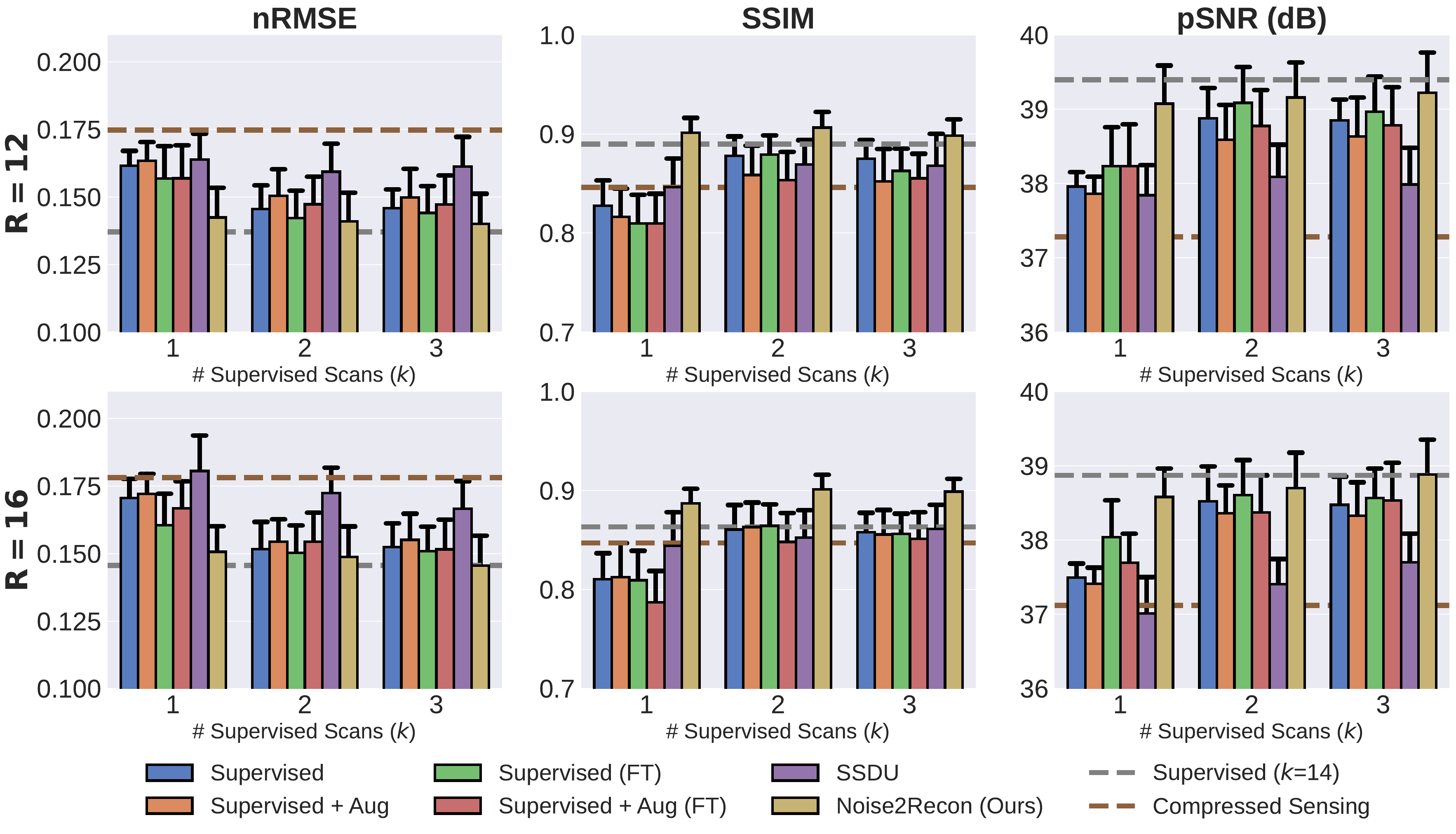}
  \end{center}
  \caption{Reconstruction performance in label scarce settings ($k=1,2,3$) at accelerations of 12x (top) and 16x (bottom). With only one supervised scan, Noise2Recon outperformed supervised methods and clinical compressed sensing baselines (brown dashed line) and approached performance of the supervised baseline trained with $k$=14 scans (gray dashed line).}
  \label{fig:label-scarcity-bar}
\end{figure}

\subsection{Baseline Comparisons}
In these experiments, we evaluate how Noise2Recon performed compared to supervised and self-supervised DL and CS baselines in (1) label-scarce settings, where only a subset of training scans have ground-truth references, and (2) OOD settings, such as low-SNR acquisitions and unseen accelerations.

\begin{figure}[t!]
  \centering
  \begin{center}
      \includegraphics[width=\linewidth]{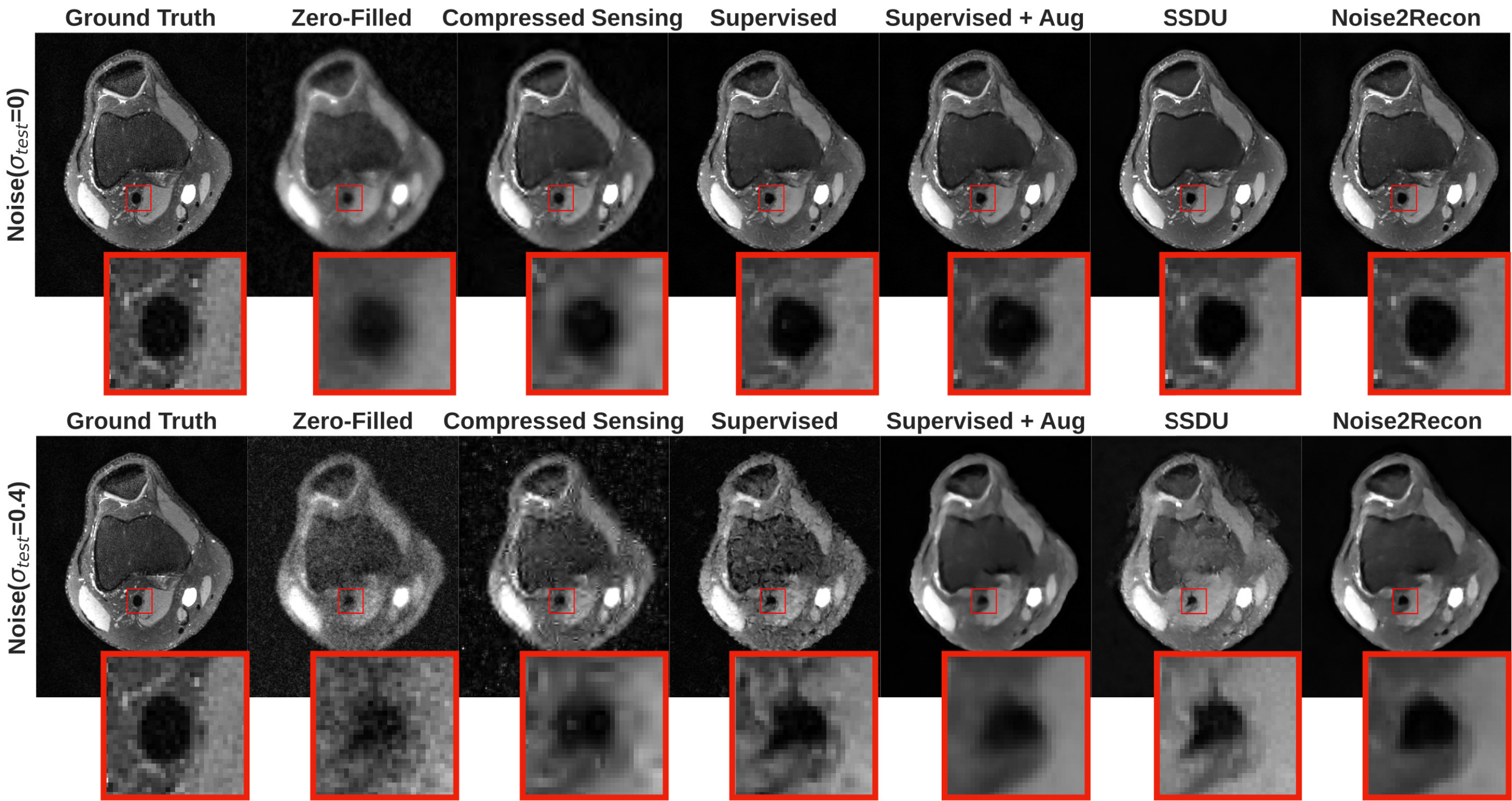}
  \end{center}
  \caption{Sample reconstructions for 12x accelerated scans in high-SNR (top) and low-SNR, out-of-distribution (bottom) settings. \textbf{Top:} With limited reference scans, Noise2Recon improves performance over supervised baselines by utilizing information from unsupervised scans through consistency regularization, resulting in high-fidelity image reconstruction. Noise2Recon preserves the morphology, sharpness, and contrast around the popliteal artery, as shown in the inset image. \textbf{Bottom:} Supervised models amplify noise artifacts, while supervised models with noise augmentations produce blurry images. Noise2Recon balances denoising and reconstruction, recovering diagnostically relevant, fine anatomical structures.}
  \label{fig:sample-images-1sub}
\end{figure}

\label{sec:results-baseline}
\paragraph{Label scarce settings} Noise2Recon outperformed both DL and CS baselines in label-scarce settings of 1 supervised scan \RV{for both feed-forward U-Net and unrolled architectures (\cref{tbl:label-scarcity-k1,tbl:label-scarcity-k1-unrolled})}. When measuring label-efficiency, Noise2Recon performed on par with supervised methods despite being trained with \textit{14 times fewer} supervised training examples (\cref{fig:label-scarcity-bar}). In addition, Noise2Recon performance did not drop as the number of supervised scans increased. Qualitatively, reconstructions with Noise2Recon had reduced blurring and noise around key anatomical structures compared to both supervised and self-supervised DL baselines trained with the same number of supervised scans (Fig. \ref{fig:sample-images-1sub}).

\begin{figure}[t!]
  \centering
  \begin{center}
      \includegraphics[width=\linewidth]{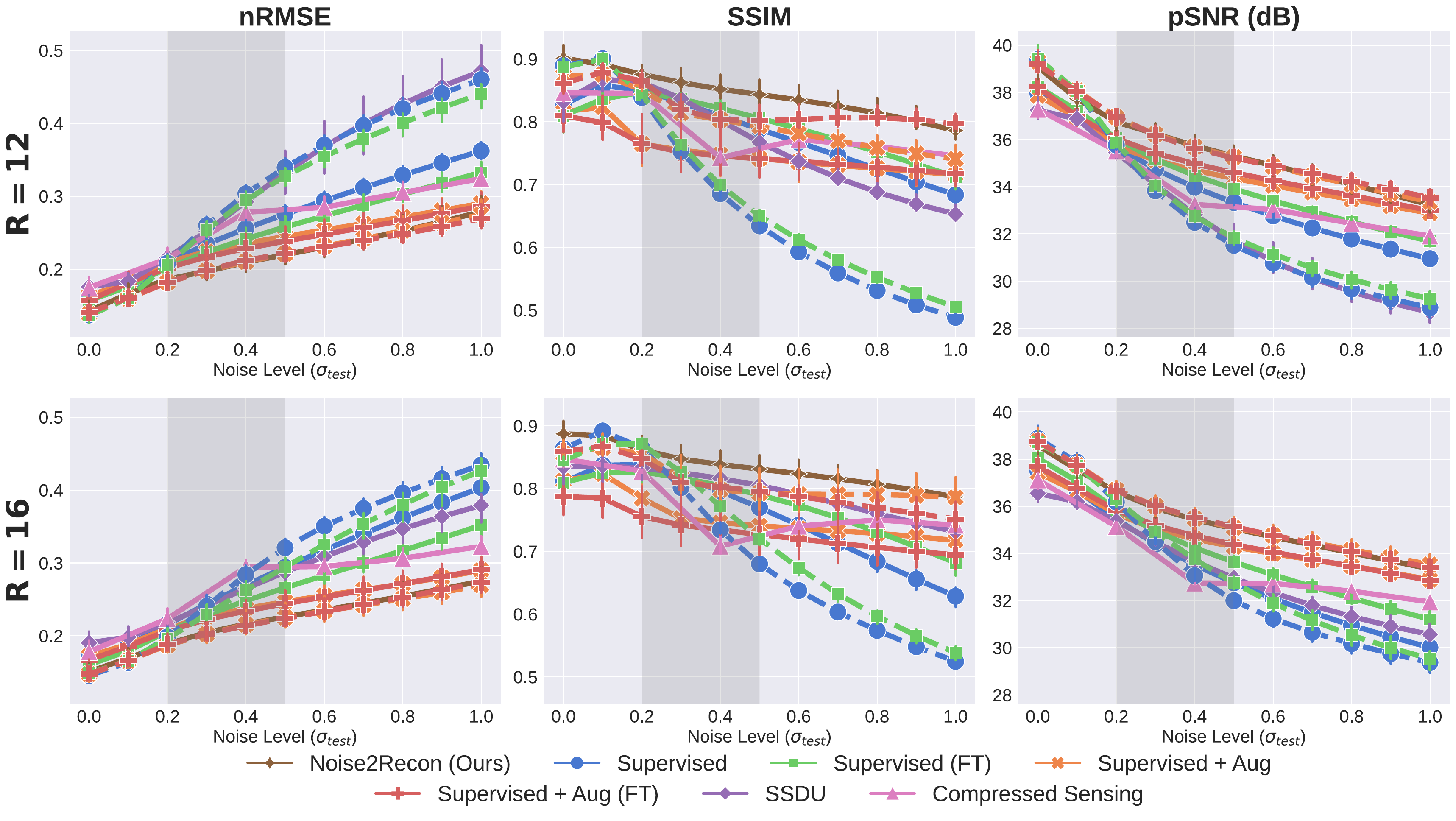}
  \end{center}
  \caption{Characterizing reconstruction performance at varying noise levels ($\testnoise>0$) at accelerations of 12x (top) and 16x (bottom). All methods were trained with $k$=1 (solid line). Supervised methods were also trained with $k$=14 (dashed line) supervised scans. Shaded area (gray) indicates training noise range ($\noiserange$). With only one supervised scan ($k$=1), \RV{Noise2Recon closes the performance gap relative to supervised methods trained with abundant supervised data ($k$=14), regardless of noise augmentations and fine-tuning}. Higher SSIM values indicate less blurring in Noise2Recon compared to CS and supervised DL methods. Noise2Recon image quality metrics had low sensitivity to increasing $\testnoise$, which may indicate higher robustness in noisy settings.}
  \label{fig:noisy-line-graph}
\end{figure}

\paragraph{Reconstructing low-SNR data} Among data-driven methods, those that used noise-based augmentations (i.e. Noise2Recon, Supervised+Aug, and Supervised+Aug-FT) achieved higher image quality compared to their non-augmented counterparts (Supervised and Supervised-FT), which amplified noise artifacts at higher noise levels (Fig. \ref{fig:noisy-line-graph}). Unlike the augmentation-based approaches, Supervised and Supervised-FT performance also deteriorated with increase in training data.
While the metrics for Supervised+Aug and Supervised+Aug-FT methods were higher than non-augmentation approaches, images reconstructed with these methods were considerably blurrier than the reconstructed images from non-augmentation baselines. In contrast, Noise2Recon sufficiently suppressed noise artifacts without excessively blurring the image (Fig. \ref{fig:sample-images-1sub}). On the other hand, Supervised resulted in amplified noise artifacts in reconstructed images, which may indicate overfitting of these methods to non-noisy scans.

\RV{Moreover, the performance of supervised augmentation baselines in both in-distribution and noisy, OOD settings was limited by the extent of fully-sampled training data (\cref{fig:noisy-line-graph})}. However, Noise2Recon recovered the performance of these Supervised+Aug and Supervised+Aug-FT models trained on the full training dataset with only one supervised training scan. Additionally, while models fine-tuned from pretrained denoisers showed improved performance in noisy settings, Noise2Recon consistently outperformed these models across all metrics. Noise2Recon also showed increased generalizability to noise levels outside of the range sampled during training ($\testnoise \notin \noiserange$) (Fig. \ref{fig:noisy-line-graph}).

\begin{figure}[t!]
    \centering
    \begin{center}
        \includegraphics[width=\linewidth]{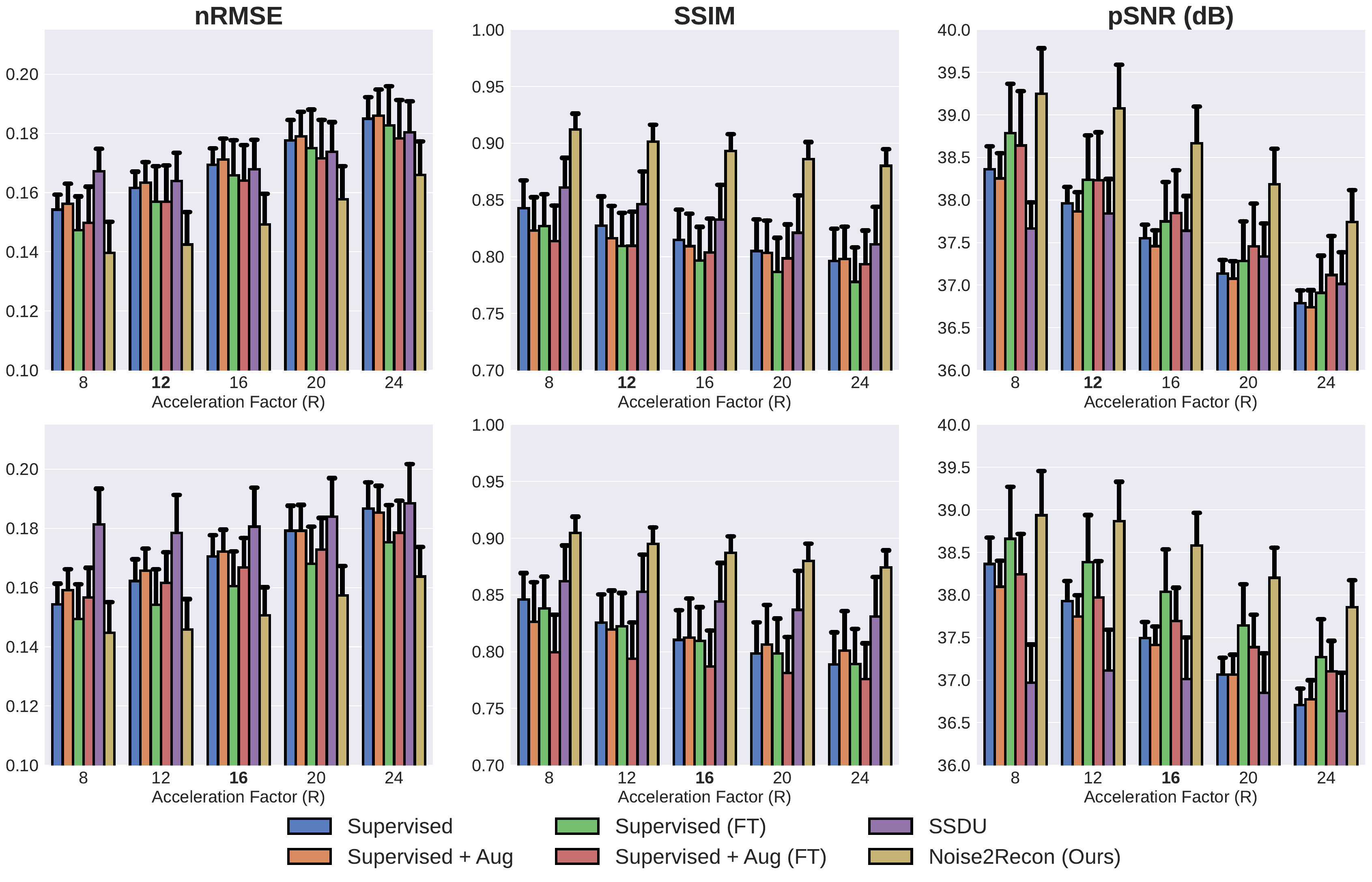}
    \end{center}
    \caption{Generalizability of methods trained on one acceleration (bolded on x-axis) to unseen accelerations. Methods were trained on scans accelerated at $R_{train}$=12 (top row) and $R_{train}$=16 (bottom row). Noise2Recon recovers images better at both lower ($R_{test}<R_{train}$) and higher acceleration factors ($R_{test}>R_{train}$) compared to supervised methods trained on the same number of supervised scans ($k=1$).}
    \label{fig:acc-one-to-many}
\end{figure}

\paragraph{Generalizing to unseen accelerations} Despite being trained on scans with a fixed acceleration factor ($R_{train}$), Noise2Recon generalized better to OOD accelerations ($R_{test} \neq R_{train}$) (Fig. \ref{fig:acc-one-to-many}). At accelerations lower than those of training scans ($R_{test}<R_{train}$), Noise2Recon reconstructions had considerably higher pSNR and SSIM than images reconstructed by supervised baselines trained with the same number of supervised scans. As the acceleration factor increased, Noise2Recon maintained higher performance than supervised methods across all metrics. Noise2Recon performance on OOD acceleration factors also surpassed that of in-distribution generalization of supervised methods. For example, at $R_{train}=12$ and $R_{test}=16$, Noise2Recon outperformed supervised methods trained on $R_{train}=16$ accelerated scans. A similar pattern was seen for $R_{train}=16$ and $R_{test}=12$.

\begin{figure}[t!]
    \centering
    \begin{center}
    \includegraphics[width=\linewidth]{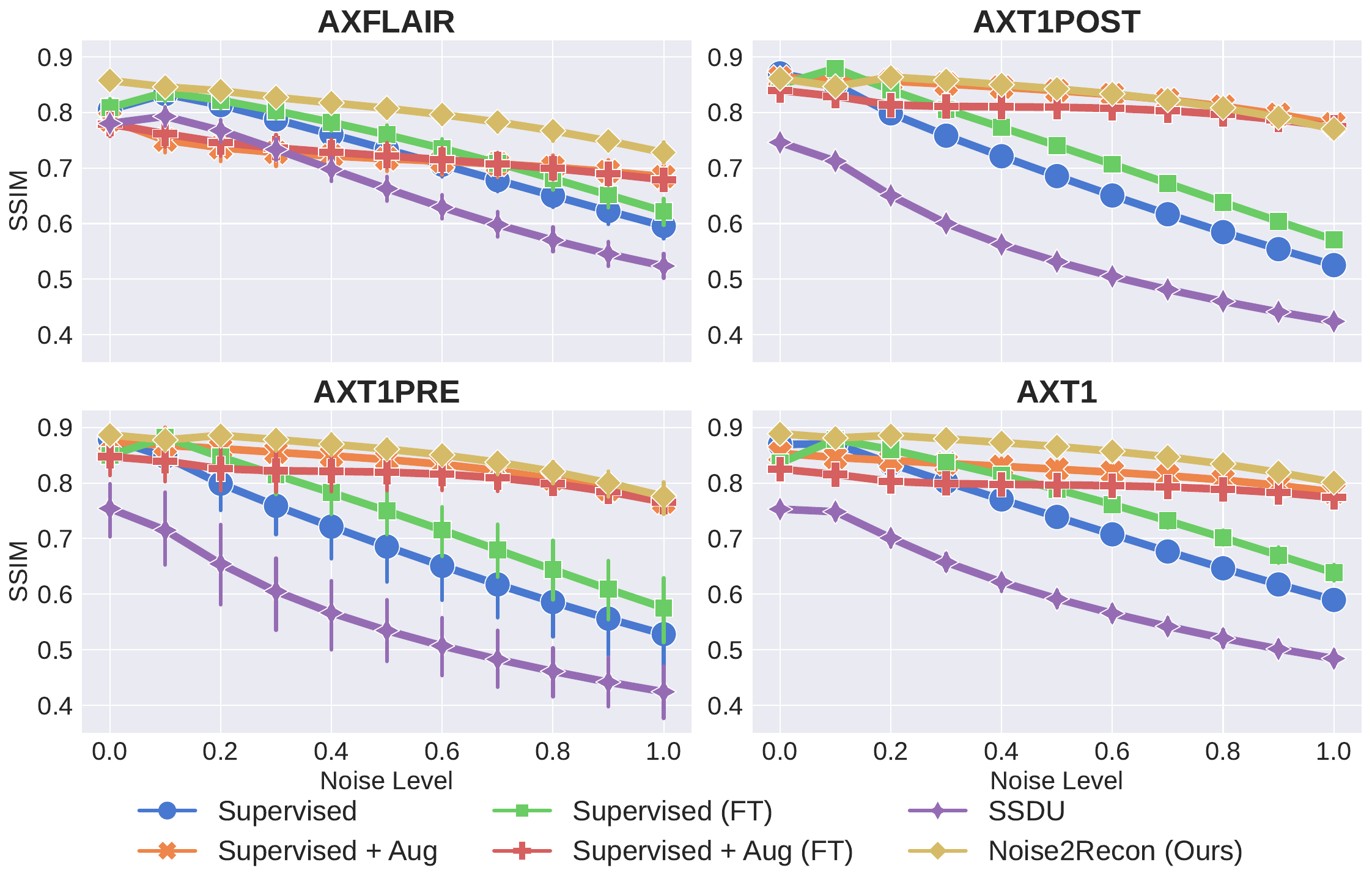}
    \end{center}
    \caption{\RV{Generalizability of U-Net models trained on 3D mridata FSE knee dataset and evaluated on 2D fastMRI brain dataset at multiple SNR levels. Noise2Recon  outperforms all baseline methods among all four acquisition types in both high-SNR ($\sigma_{test}=0$) and low-SNR ($\sigma_{test}>0$) settings.}}
    \label{fig:fastmri-unet-noise}
\end{figure}

\begin{figure}[t!]
    \centering
    \begin{center}
        \includegraphics[width=\linewidth]{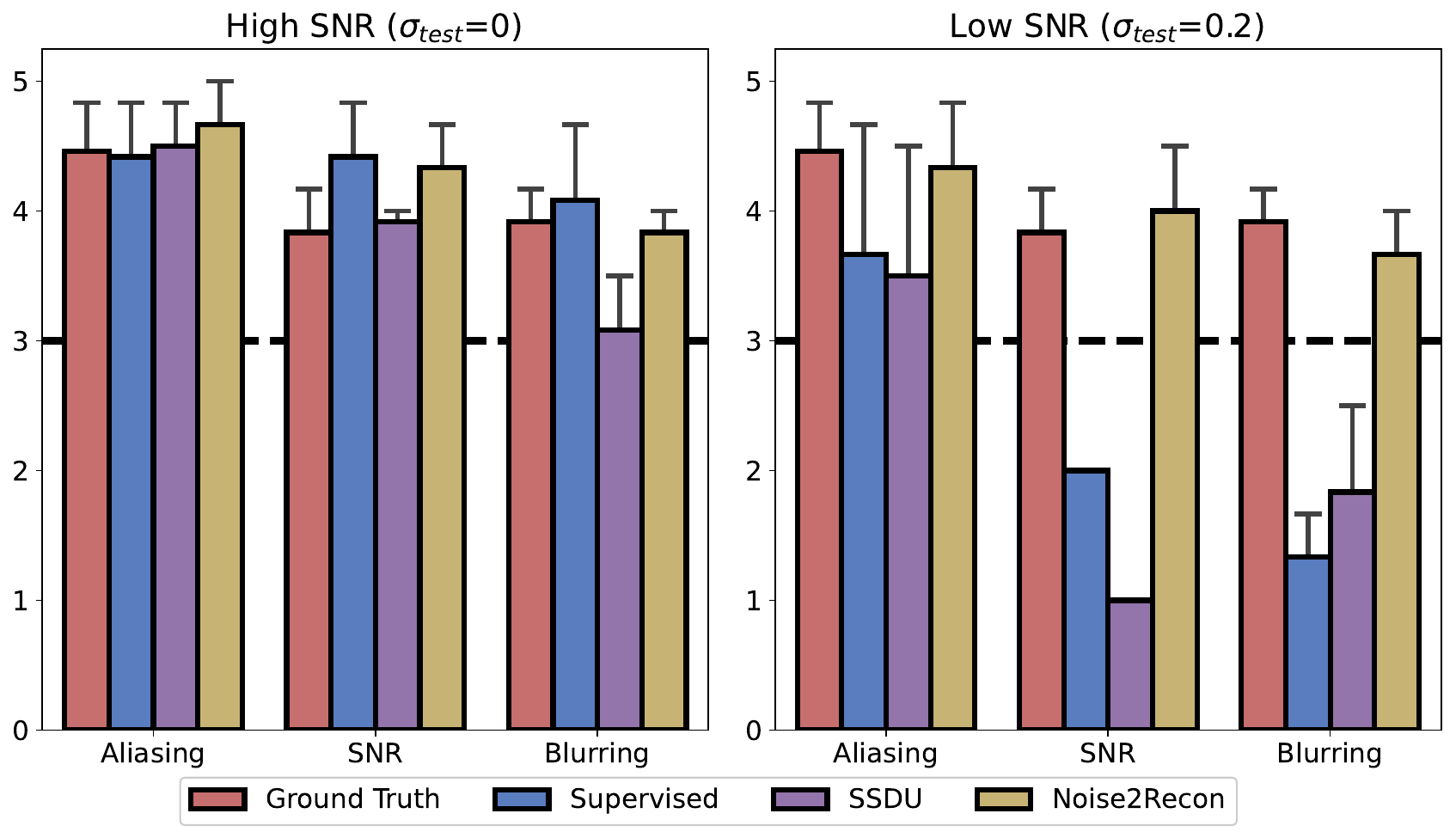}
    \end{center}
    \caption{\RV{Results (mean $\pm$ std. dev.) from the radiologist reader study. Methods were compared in both high-SNR, in-distribution (left) and low-SNR, out-of-distribution (right) settings. In high-SNR settings, all DL reconstructions have similar or slightly better aliasing, SNR and blurring artifacts compared to ground truth reconstructions. In low-SNR settings, Noise2Recon has considerably better performance across all artifacts compared to baseline self-supervised (SSDU) and supervised methods. Noise2Recon also recovers images with comparable SNR and aliasing quality to ground truth references. In both settings, Noise2Recon reconstruction quality was above the minimum diagnostic quality (dashed line).}}
    \label{fig:reader-study}
\end{figure}

\begin{figure}[t!]
    \centering
    \begin{center}
    \includegraphics[width=\linewidth]{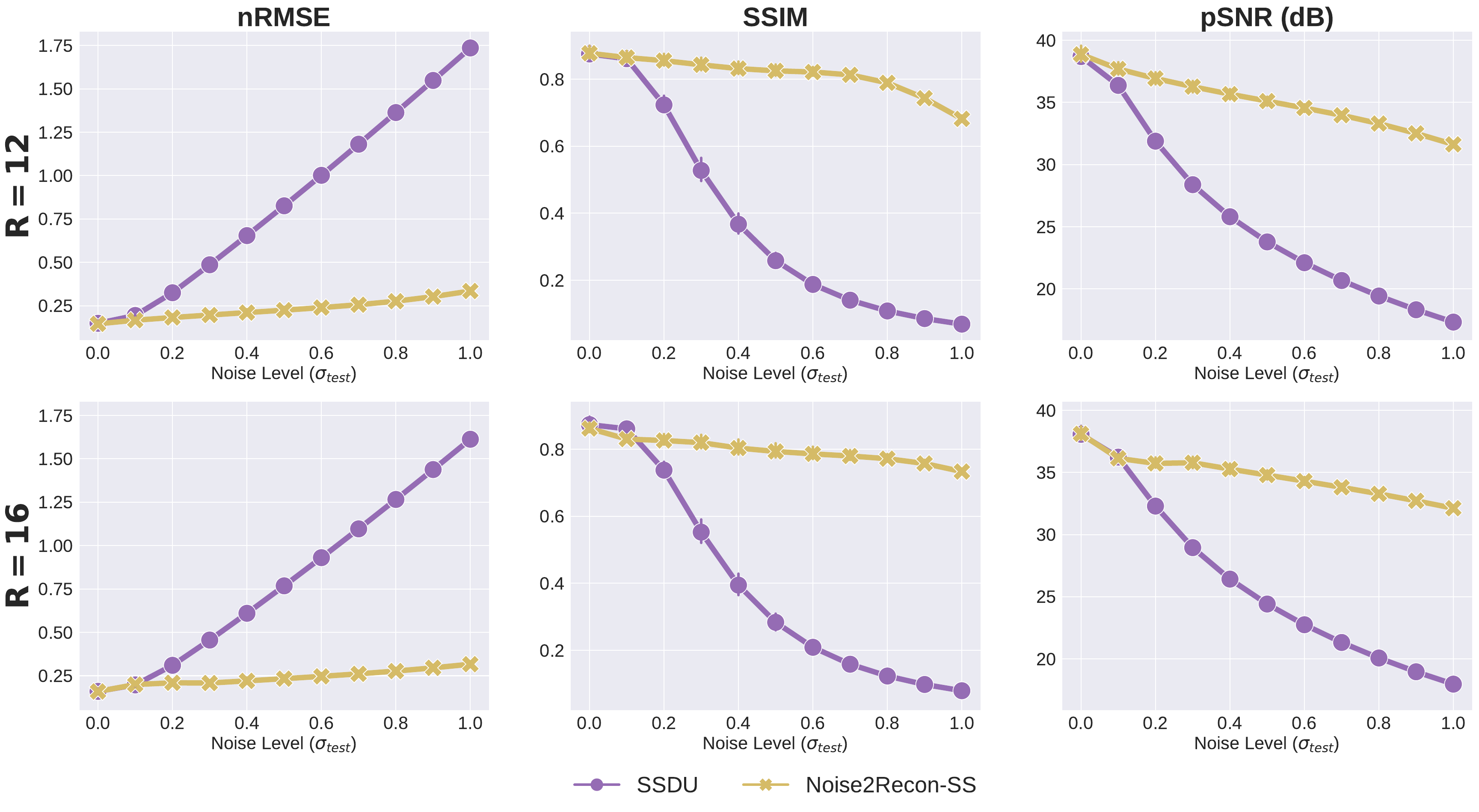}
    \end{center}
    \caption{\RV{Unsupervised methods. The self-supervised extension of Noise2Recon (Noise2Recon-SS) and SSDU perform comparably on high-SNR data ($\sigma_{test}=0$). While SSDU performance degrades with increasing noise ($\uparrow \sigma_{test}$), Noise2Recon is more robust to changes in SNR.}}
    \label{fig:unsupervised}
\end{figure}

\RV{
\paragraph{Cross-dataset generalizability} Noise2Recon also generalized to \textit{both high-SNR and low-SNR settings} in the 2D fastMRI brain dataset. At high-SNR ($\sigma_{test}=0$), Noise2Recon outperformed all baseline methods with the U-Net architecture (\cref{fig:fastmri-unet-noise}) and performed comparably to SSDU with unrolled networks (\cref{fig:fastmri-unrolled-noise}). Among challenging low-SNR scans ($\sigma_{test}>0$), Noise2Recon achieved better performance compared to all other baselines (\cref{fig:fastmri-unet-noise}).

\paragraph{Reader study} Noise2Recon had similar radiologist-evaluated perceptual scores to ground truth reference reconstructions in terms of aliasing, SNR, and blurring artifacts (\cref{fig:reader-study}). In low-SNR settings, Noise2Recon outperformed SSDU and supervised methods across all artifacts.

\paragraph{Unsupervised settings}  Noise2Recon-SS, the self-supervised variant of our method that does not require any labeled data, achieves comparable performance to SSDU in high-SNR settings and considerably outperforms SSDU among low-SNR scans (\cref{fig:unsupervised}).
}

\subsection{Ablation Study}
\label{sec:results-ablation}
In these experiments, we investigate three natural design questions that may be helpful for training Noise2Recon:
\begin{enumerate}
    \item How should supervised and unsupervised data be sampled during training?
    \item How should the training noise range ($\noiserange$) for training augmentations be configured?
    \item How should loss weighting be selected?
\end{enumerate}
We show that Noise2Recon is not very sensitive to any of these design decisions (especially 2\&3), which may reduce the burden of hyperparameter search during training. All ablations are performed on $k=1$ configurations with the same hyperparameters detailed in \cref{app:additional-exps}.

\paragraph{Balanced data sampling} We evaluated the impact of balanced sampling between supervised ($S$) and unsupervised ($U$) examples during training. \cref{fig:abl-balanced-sampling} shows the performance of balanced sampling with different $T_S$:$T_U$ ratios compared to random sampling. Regardless of the ratio, balanced sampling consistently outperforms random sampling across all metrics. Oversampling supervised scans relative the unsupervised scans ($T_S>T_U$) performed slightly better than oversampling unsupervised scans ($T_U>T_S$). The top two overall performances across all metrics were achieved with $T_S$:$T_U$ ratios of 2:1 and 1:1, respectively.

\paragraph{Sensitivity to training noise levels} We consider two training techniques that may impact the overall difficulty of learning to generalize from augmentations: noise ranges $\noiserange$ (1) with larger intervals that increase variance of sampled noise augmentations, and (2) with larger upper bounds that account for a higher magnitude of noise-corruption. The performance of Supervised+Aug models deteriorated more rapidly with increased noise, particularly among metrics emphasizing high-frequency information such as SSIM (\cref{fig:noise-lvl-ablation-multi-noise}E). Meanwhile, Noise2Recon generalized better to both in-distribution scans ($\testnoise=0$) and OOD, noisier scans (\cref{fig:noise-lvl-ablation-multi-noise}). All networks trained with small noise intervals (i.e. $\noiserange$ is small) did not generalize at higher noise settings. For Noise2Recon, this was mitigated by either increasing the upper bound of the noise range or increasing the size of the noise range.

\paragraph{Sensitivity to loss weighting} We investigated the impact of the consistency loss weighting parameter $\lambda$ on overall performance of Noise2Recon models. In the in-distribution evaluation setting, the weighting factor had negligible impact on performance between $\lambda \in [0.05, 0.8)$ (Fig. \ref{fig:ablation-cons-loss-weighting}). At very low ($\lambda \leq 0.01$) or high ($\lambda \geq 0.8$) weighting factors, metrics reduced slightly, but within the error range. Among simulated noisy acquisitions, Noise2Recon reconstruction performance for $\lambda \in [0.05, 0.8)$ was also similar for all testing noise levels $\testnoise \in \{0, 0.1, \dots, 1\}$.

\section{Discussion}
In this work, we propose Noise2Recon, \RV{a model-agnostic, label-efficient approach} for joint MRI reconstruction and denoising that can leverage both fully-sampled and undersampled scans to (1) minimize dependence on supervised data and (2) improve reconstruction performance in various OOD settings, such as low-SNR, changes in acceleration, and dataset shifts. We show that augmentation-based consistency is a viable method for recovering performance in label-scarce and OOD settings compared to CS and supervised and self-supervised DL methods. \RV{We also demonstrate self-supervised Noise2Recon (Noise2Recon-SS) is effective in unsupervised settings, where fully-sampled data is unavailable.} In this section, we first explore the relationship between our method and principles in both compressed sensing and \RV{multi-objective} learning. We then discuss the practical utility of our method in label-limited and OOD settings. Finally, we detail characteristics of our method that can improve network optimization and simplify training.

\paragraph{Learning in label-scarce settings} As model performance is proportional to the size and quality of the training data and labels \cite{hestness2017deep,ratner2016data}, standard supervised methods often fail in label-limited regimes. Noise2Recon enables joint usage of supervised data and unsupervised data to complement the image reconstruction task while increasing robustness when reconstructing noisy scans. The improved performance of Noise2Recon over supervised baselines among in-distribution scans may indicate that consistency training can (1) improve the estimation of the true data distribution with more training examples and (2) generate high quality pseudo-labels that can function as noisy surrogates for the true labels without impairing training from supervised examples.

\paragraph{Robustness to noise} Differences in the SNR among MR acquisitions are pervasive given the heterogeneity of MR hardware (e.g. field strength, coil geometry) and sequence parameters (e.g. echo time). Reconstruction methods that can generalize better to such distribution shifts may have practical utility for prospective deployment. \RV{With minimal or no fully-sampled scans, Noise2Recon improved performance among low-SNR scans without impairing performance on in-distribution, high-SNR ($\testnoise=0$) examples.} Noise2Recon generalized across all testing noise levels and improved visual quality, which may indicate that Noise2Recon simultaneously minimizes global error (i.e. mean squared error) \textit{and} recovers fine anatomical structure.  Thus, Noise2Recon (1) demonstrates utility for label-efficiency in addressing distribution shift settings and (2) can generalize better to acquisitions at different noise levels, even compared to supervised methods with ample training data, without collapsing towards the trivial denoising solution (i.e. blurring).



\paragraph{Generalization under unseen distribution shift} It is intractable to capture training data for the exhaustive set of acquisition settings in which the model should perform well. As such, it is practically useful for DL methods to be able to generalize to perturbations that were not simulated during training (i.e. \textit{unseen} settings). Noise2Recon generalized to unseen noise levels ($\testnoise \notin \noiserange$), unseen accelerations ($R_{test} \neq R_{train}$), and even \RV{compounding OOD factors (e.g. sampling pattern, acquisition, field strength, etc.) often found in dataset shifts}. The improved performance in these settings may suggest that the joint optimization of the reconstruction and denoising objectives contributes to positive transfer between the two tasks \cite{wu2020understanding}. This observation may empirically validate that even among DL methods, noise is a reasonable model for signal incoherence, as is proposed in CS theory \cite{Lustig_Donoho_Pauly_2007}. Thus, learning to denoise images can also help improve reconstruction efficacy in cases where aliasing is extensive (i.e. higher accelerations). Overall, Noise2Recon may be more robust in response to larger extents of distribution shift than supervised DL methods and may be a more viable candidate for deployment in different acquisition settings.


\paragraph{Stabilizing multi-objective optimization} As mentioned in \S\ref{sec:method-n2r}, the magnitude of the supervised and consistency objectives are implicitly weighted by the number of supervised and unsupervised examples. Thus, random data sampling may lead to sub-optimal convergence for both objectives. Balanced sampling can eliminate this weighting factor by controlling the duty cycle of supervised and unsupervised examples during optimization. We find that this sampling procedure can improve overall performance. This technique is also reminiscent of sub-group sampling in methods in distributionally robust optimization for classification models, where examples of classes are sampled at a rate inversely proportional to the class frequency \cite{sagawa2019distributionally}.

\paragraph{Insensitivity to hyperparameter selection} Multi-objective training frameworks and augmentation optimization often require careful hyperparameter tuning due to optimization instabilities introduced with different simulated data distributions or weighted objectives. However, Noise2Recon showed minimal sensitivity to hyperparameter selection, specifically the training noise range $\noiserange$ and the consistency loss weighting $\lambda$. Training noise ranges in Noise2Recon could be increased without degrading performance across any noise levels. Additionally, Noise2Recon was generally insensitive to a wide range of weighting parameters $\lambda$, in contrast to most multi-objective methods that require tuning for superior performance. This may suggest that the consistency training in Noise2Recon can minimize instabilities in network optimization caused by small changes in hyperparameters and may be practically useful for simplifying network training.

\paragraph{Limitations and future work}
While the the scope of this study was limited to noise augmentations, the consistency regularization paradigm used in Noise2Recon may be extendable to other artifacts observed in MRI such as motion, $B_0$ inhomogeneity, phase wrapping, and eddy currents. \RV{Additionally, augmentations in Noise2Recon can be combined with curriculum learning \cite{braun2017curriculum} and minimax augmentation sampling methods \cite{gnanasambandam2020one} to increase generalizability to large, OOD noise settings. Moreover, Noise2Recon demonstrated high performance with simulated SNR changes. In future work, we will investigate how Noise2Recon generalizes to prospectively accelerated, low-SNR settings (e.g. lower field strength, different coils).}

\section{Conclusion}
In this work, we propose Noise2Recon, a label-efficient, consistency-based approach for joint MRI reconstruction and denoising. We demonstrated that Noise2Recon can outperform standard supervised methods in both in-distribution and \RV{OOD settings (e.g. low-SNR, acceleration shift, and cross-dataset}, with limited training data. \RV{In addition, we showed Noise2Recon can be extended to both semi-supervised and self-supervised settings}. By reducing dependence on supervised data for model training and increasing generalizability to various OOD factors, Noise2Recon shows potential for reducing the burden of model retraining or fine-tuning in both research and clinical settings.

\section{Acknowledgements}
This work was supported by R01 AR063643, R01 EB002524, R01 EB009690, R01 EB026136, K24 AR062068, and P41 EB015891 from the NIH; the Precision Health and Integrated Diagnostics Seed Grant from Stanford University; DOD – National Science and Engineering Graduate Fellowship (ARO); National Science Foundation (GRFP-DGE 1656518, CCF1763315, CCF1563078); Stanford Artificial Intelligence in Medicine and Imaging GCP grant; Stanford Human-Centered Artificial Intelligence GCP grant; GE Healthcare and Philips.

\bibliographystyle{plainnat}
\bibliography{ms}

\clearpage

\begin{table*}[t]

\centering
    \caption{\small Summary of notation used in this work.}
    \label{tab:glossary}
    \resizebox{\textwidth}{!}{%
    \begin{tabular}{@{}lll@{}}
      \toprule
      & Notation                                            & Description                                                                                           \\
      \midrule
      \textbf{MRI forward model} & $x, y$                    & Image, k-space measurements  \\
      & $x^{*}, y^{*}$                                             & True image, k-space \\
      & $\hat{x}, \hat{y}$                                         & Predicted image, k-space \\
      & $\Omega, F, S$                                           & Undersampling mask, \RV{Fourier} transform matrix, coil sensitivity maps                                            \\
      & $A$                                                     & Forward MRI acquisition operator \\
      & $\epsilon$                                                 & Additive complex-valued Gaussian noise                                                              \\
      
      \textbf{Data} & $\mathcal{D}^{(s)}, \mathcal{D}^{(u)}$ & \makecell[l]{Dataset of fully-sampled (i.e. supervised, labeled) scans, \\ prospectively undersampled (i.e. unsupervised, unlabeled) scans} \\
      & $\mathcal{D}$ & The total dataset (i.e. $\mathcal{D}^{(s)} \cup \mathcal{D}^{(s)}$) \\
      & $\mathcal{D}_k$ & Dataset with $k$ fully-sampled examples and $|\mathcal{D}|-k$ undersampled examples \\
      & $y^{(s)}_i, y^{(u)}_j$                  &  \makecell[l]{K-space of fully-sampled example, prospectively undersampled example \\ where $y^{(s)}_i \in \mathcal{D}^{(s)}, y^{(u)}_j \in \mathcal{D}^{(u)}$} \\
      & $\Omega_{y^{(u)}_j}$ & Undersampling mask for example $y^{(u)}_j$ \\
      & $R_{train}, R_{test}$ & Acceleration used for training, testing \\
      
      \textbf{Noise Augmentation} & $\mathcal{N}(\mu, \sigma)$ & Gaussian (normal) distribution with mean $\mu$ and standard deviation $\sigma$ \\
      & $\epsilon_j$ & Simulated noise map for undersampled example $y^{(u)}_j$ \\
      & \RV{$\tilde{\epsilon}_j$} & \RV{Masked noise map (in Fourier domain) for undersampled example $y^{(u)}_j$} \\
      & $\noiserange$ & Range of standard deviations for noise augmentations \\
      & $\sigma_{tr}^L, \sigma_{tr}^U$ & Lower, upper bounds for $\noiserange$ \\
      & $\testnoise$ & Noise standard deviation used for testing \\
      & $p$ & Augmentation probability \\

      \textbf{Model components} & $f_{\theta}$ & The model parametrized by $\theta$\\ 
      \textbf{and losses}   & $\mathcal{L}_{sup}$, $\mathcal{L}_{cons}$      & Supervised, consistency loss\\
          &     $\lambda$      & Consistency loss weight \\
          & \RV{$\mathbb{E}$} & \RV{The expectation of a random variable} \\
         & $T_S$, $T_U$ & Supervised, unsupervised periods for balanced sampling \\
      \bottomrule
    \end{tabular}%
    }
\end{table*}

\section*{Appendix}
\appendix

\section{Glossary}
\label{app:glossary}
\cref{tab:glossary} provides a summary of the notation used in the paper.
\section{Experimental Details}
This section describes experiment details for Noise2Recon and baselines. All code, experimental configurations, and pre-trained models are available at \url{https://github.com/ad12/meddlr}.

\subsection{Baselines}
\subsubsection{Pretrained Denoisers \& Fine-Tuning}
\label{app:pretrained-denoisers-implementation}
\RV{To investigate the efficacy of using denoising networks for MRI reconstruction, we compared Noise2Recon to a family of baselines where pretrained denoisers were fine-tuned on the MRI reconstruction task. This baseline had a two-stage training protocol: 1) self-supervised denoising pretraining and 2) supervised MRI reconstruction fine-tuning.


\paragraph{Denoising pretraining.} Denoising networks were trained mridata 3D FSE knee training dataset following the protocol proposed in Noise2Self \cite{batson2019noise2self}. Because denoising can be formulated as a self-supervised problem, the model was trained with both fully-sampled and prospectively undersampled data. As a source of data augmentation, fully-sampled scans were undersampled following the same undersampling pattern (Poisson Disc) and acceleration rate that would be used during fine-tuning. All examples were augmented with zero-mean complex-Gaussian masked noise with standard deviation $\sigma_{tr}$ sampled from range $\noiserange$. The model was trained to recover the original, non-augmented image from the noise-augmented input. We refer to the output of this stage as the \textit{pretrained model}. 

\paragraph{MRI reconstruction fine-tuning.} The pretrained model was subsequently fine-tuned on the MRI reconstruction task using only fully-sampled data (i.e. supervised training). Two supervised training protocols were followed - training without any noise augmentations (i.e. \textit{Supervised-FT}) and with noise augmentations (i.e. \textit{Supervised+Aug-FT}).}

\subsubsection{Self-supervised Learning via Data Undersampling (SSDU)}
\label{app:ssdu}
SSDU was originally proposed for fully unsupervised settings, where all training data are prospectively undersampled. For fair comparison to Noise2Recon, which is a semi-supervised method, we propose a trivial extension to adapt SSDU to the semi-supervised setting. For prospectively undersampled (unsupervised) scans, the training strategy proposed in SSDU was used. Examples sampled from the fully-sampled (supervised) training set were retrospectively undersampled using a random undersampling mask generated from the undersampling method and acceleration factor for the given experiment. These simulated undersampled scans were used as inputs to the SSDU protocol. The random undersampling mask was generated dynamically -- i.e. each time a fully-sampled example was sampled for training, a unique undersampling mask was used. This protocol serves as a method of augmentation for supervised scans.

\begin{figure}[t!]
  \centering
  \begin{center}
      \includegraphics[width=\linewidth]{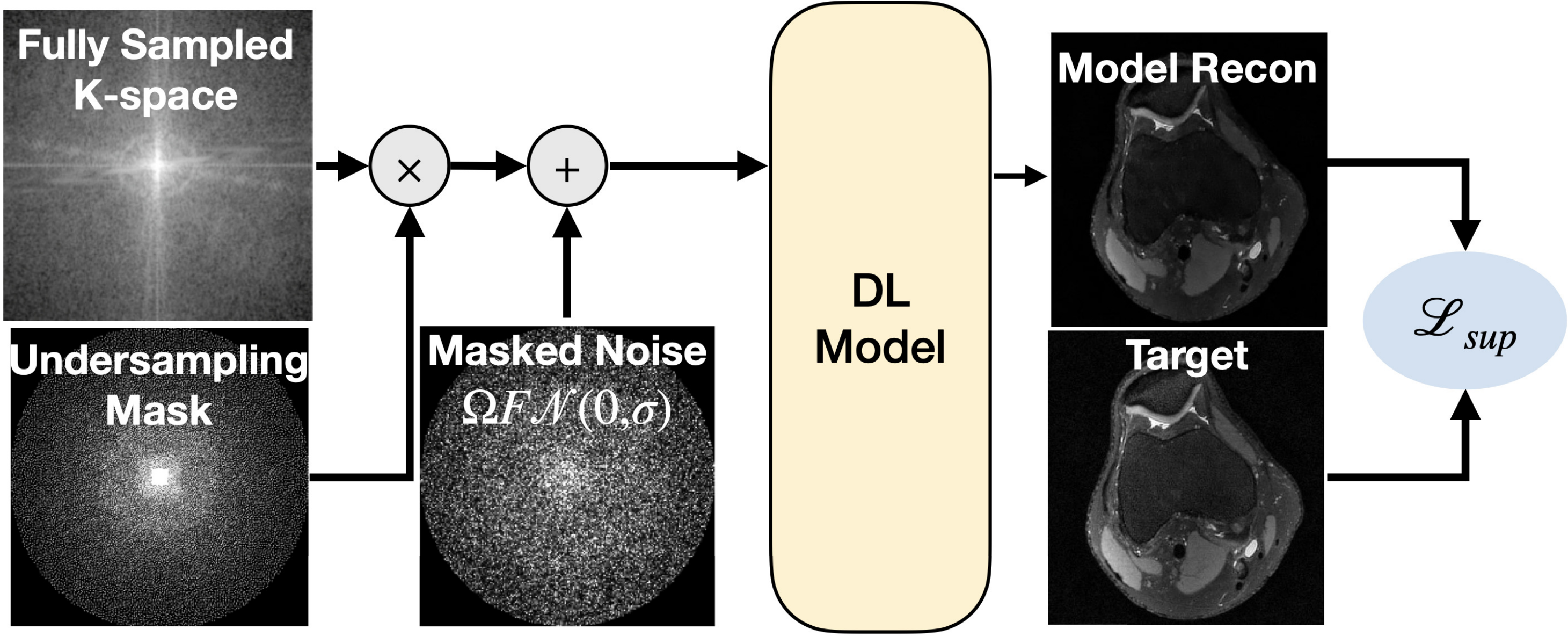}
  \end{center}
  \caption{Example of k-space noise augmentations used in supervised training. Fully-sampled scans are retrospectively undersampled and corrupted with masked additive noise. The noisy, undersampled k-space is reconstructed by the model and compared to the target image, which is computed by applying the forward acquisition operator $A$ to the fully-sampled k-space.}
  \label{fig:supervised-aug-schematic}
\end{figure}

\RV{\paragraph{Postprocessing: hard data consistency} We find that SSDU networks are sensitive to the use of data consistency (DC). However, standalone feed-forward CNNs, like U-Net, do not have data consistency by definition. Additionally, hard data consistency post-processing (e.g. \cite{darestani2021accelerated}) fails when the observed k-space samples are corrupted (e.g. low-SNR acquisitions). To address these issues among feed-forward networks, we propose a variant of hard DC termed \textit{edge hard DC}, where edge regions of the reconstructed k-space $\hat{y}$ are replaced with the edge regions of the acquired k-space $y$. The "edge region" is defined as the outer regions of the acquired k-space with no signal. More formally, given a mask $\Theta$, which is 1 for all edge locations in k-space, the postprocessed k-space $\hat{y}_{pp}$ can be written as follows:

\begin{equation*}
    \hat{y}_{pp}[i,j] = \begin{cases}
        y[i,j] & \text{if } \Theta[i,j]=1\\
        \hat{y}[i,j] & \text{if } \Theta[i,j]=0\\
    \end{cases}
\end{equation*}
}

\subsubsection{Compressed Sensing (CS)}
\label{app:compressed-sensing-implementation} 
In CS, careful tuning of the regularization parameter $\lambda$ is required for each application. As the noise level $\sigma$ and the acceleration factor $R$ are varied, the sparsity level and the blurring of the input zero-filled image changes. Therefore, the regularization parameter $\lambda$ was chosen based on visual tuning for various noise levels and acceleration factors independently. We observed that a high $\lambda$ is needed at high noise levels $\sigma$ to preserve reconstruction fidelity, whereas a lower $\lambda$ is needed at lower acceleration factors $R$ to prevent blurring. In contrast to CS, Noise2Recon does not require visual tuning of its parameters, and is more robust to different noise levels at inference time.


\subsection{Hyperparameters}
\label{app:hyperparameters}
 
\paragraph{U-Net Architecture and Optimization.} 2D U-Net models \cite{ronneberger2015u} were configured with 4 pooling layers, where the first convolution in the model had 32 output channels. Each resolution of the U-Net consisted of a convolutional block with two $3 \times 3$ convolutions followed by instance normalization and a leaky Rectified Linear Unit (ReLU) with slope $\alpha$=-0.2. \RV{The model had 7.76M trainable parameters.} Models were trained for 80,000 iterations \RV{($\sim$286 epochs relative to full training dataset)} with the Adam optimizer \cite{kingma2014adam} with default parameters ($\beta_1$=0.9, $\beta_2$=0.999), weight decay 1e-4, learning rate $\eta$=1e-3, and batch size of 16. U-Net models were trained with the complex $\ell_1$ image loss, unless otherwise specified.\\

\paragraph{Unrolled Architecture and Optimization.} Unrolled networks followed the fast iterative shrinkage-thresholding algorithm (FISTA) unrolled architecture \cite{xin2022fista} implemented in \cite{sandino2020compressed}. The network consisted of 8 unrolled blocks, where each block consisted of two residual sub-blocks. The model had 3.58M trainable parameters. All models were trained for 80,000 iterations ($\sim$38 epochs relative to full training dataset) with the Adam optimizer \cite{kingma2014adam} with default parameters ($\beta_1$=0.9, $\beta_2$=0.999), weight decay 1e-4, and learning rate $\eta$=1e-4. Given memory constraints, a batch size of 2 with 8x gradient accumulation was used to achieve an effective batch size of 16. All models were trained with the complex $\ell_1$ k-space loss, unless otherwise specified.

\paragraph{Supervised Training with Augmentations.} To tune the probability of applying augmentations, a hyperparameter sweep was conducted for probabilities $p=0.1,0.2,0.3,0.5$. The configuration with the lowest validation loss ($p$=0.2) was selected for all experiments. By default, the training noise range of $\noiserange=[0.2,0.5)$ was used.

\paragraph{Denoising Pretraining \& Supervised Fine-Tuning.} For comparison, number of training steps was split evenly between denoising pretraining and reconstruction fine-tuning. Denoising pretraining was performed for half of the total length of training (i.e. 40,000 iterations). The training noise range $\noiserange=[0.2, 0.5)$ was chosen to be consistent with the range used for Noise2Recon and Supervised+Aug methods. All denoisers were trained with the complex-$\ell_1$ objective. For fine-tuning, the network was initialized with the weights resulting in the lowest validation loss during the denoising and trained following the supervised protocol detailed in \S\ref{sec:methods-baseline}. 

\paragraph{SSDU.} SSDU can be sensitive to both the loss function and masking extent $\rho$. \RV{We perform a binary grid search for both the loss function and masking extent $\rho$.} For the U-Net model, we use the configuration recommended by \cite{Yaman_self} with the normalized k-space $\ell_1-\ell_2$ loss and $\rho$=0.4 masking extent. \RV{For the unrolled network, we use the complex $\ell_1$ k-space loss and  $\rho$=0.2.}

\paragraph{Compressed Sensing.} As discussed in \cref{app:compressed-sensing-implementation}, the regularization weight $\lambda$ must be tuned. The optimal $\lambda$ found for each noise level is provided in Table \ref{tbl:cs-reg}. Independent of $\lambda$, CS converged at 100 iterations, where we observed that higher number of iterations did not improve performance.

\begin{figure}[t!]
  \centering
  \begin{center}
      \includegraphics[width=\linewidth]{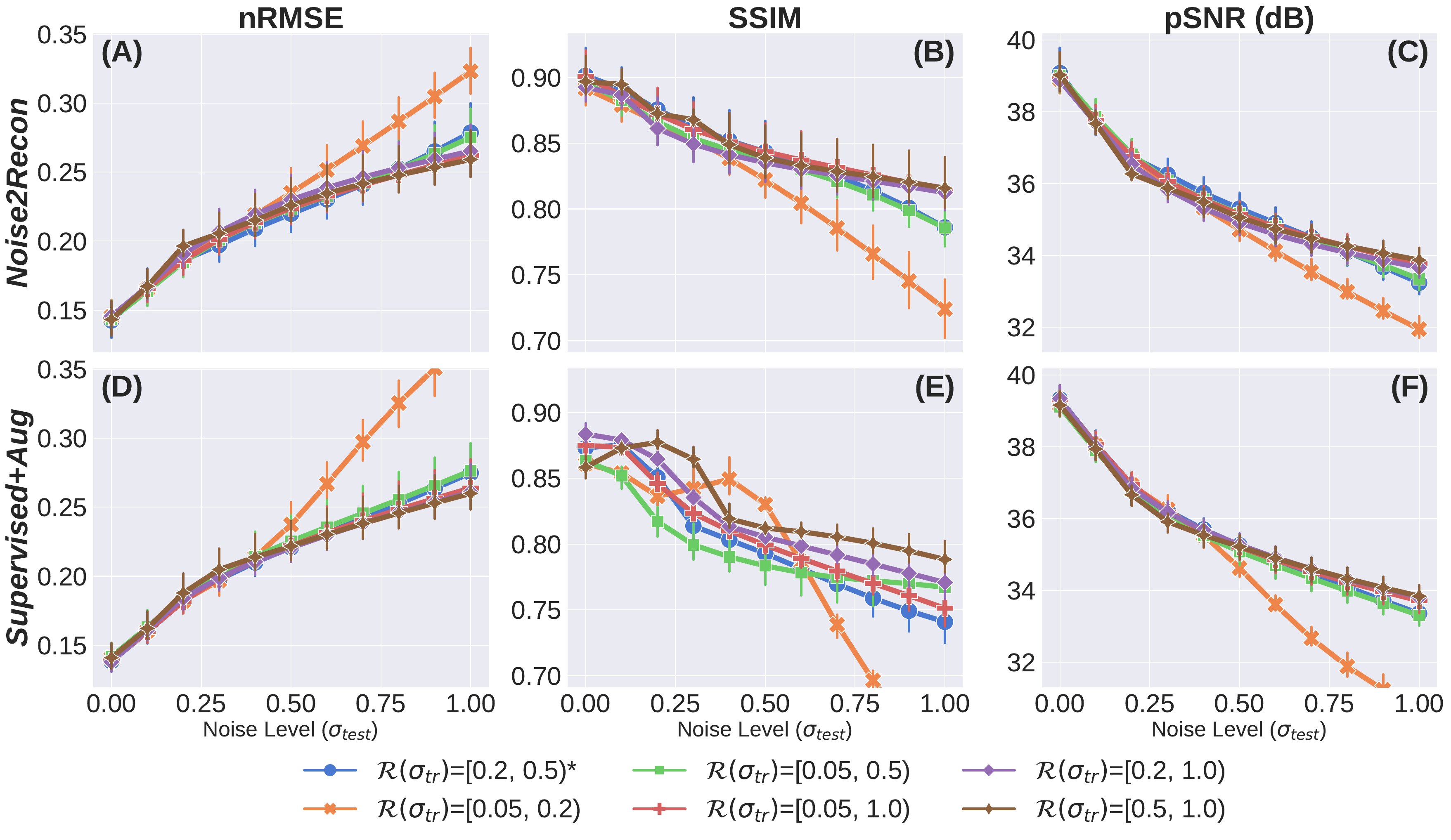}
  \end{center}
  \caption{Impact of training noise level $\sigma_{\text{tr}}$ among Noise2Recon trained with 1 supervised scan (A-C) and supervised models trained with 14 supervised scans (D-F). Performance is measured at multiple testing noise levels. Asterisk (*) indicates the default training noise level range for experiments. Noise2Recon is less sensitive to changes in $\sigma_{\text{tr}}$ compared to supervised methods with noise augmentations. Higher SSIM in Noise2Recon were consistent with considerably less blurring compared to supervised methods.}
  \label{fig:noise-lvl-ablation-multi-noise}
\end{figure}

\begin{figure}[t!]
  \centering
  \begin{center}
      \includegraphics[width=\linewidth]{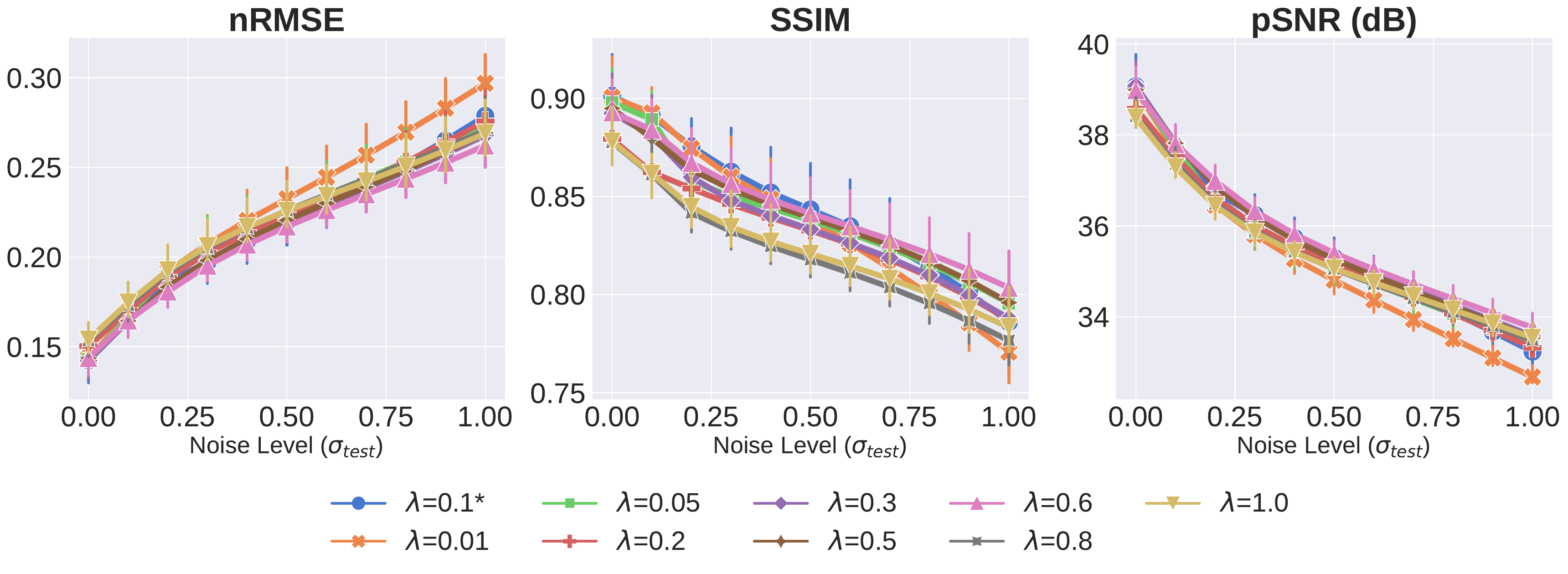}
  \end{center}
  \caption{Impact of consistency loss weighting $\lambda$ on reconstructing scans at different noise levels. Asterisk (*) indicates the default loss weighting configuration for experiments. Performance of Noise2Recon did not change for large range of $\lambda \in [0.05, 0.8)$. Insensitivity to changes in $\lambda$ may help eliminate the need for hyperparameter tuning, which can simplify network training.}
  \label{fig:ablation-cons-loss-weighting}
\end{figure}

\paragraph{Noise2Recon.} Noise2Recon models used for comparison with baseline methods were trained with 1:1 balanced sampling and training noise range of $\noiserange=[0.2,0.5)$ in concordance with all other noise augmentation baselines. Consistency loss weight was $\lambda=0.1$. We note that none of these hyperparameters were actively tuned for Noise2Recon. In fact, the training noise range $\noiserange$ and $\lambda$ used in these experiments are not the best performing parameter configuration (\cref{fig:noise-lvl-ablation-multi-noise,fig:ablation-cons-loss-weighting}). To demonstrate that Noise2Recon is less sensitive to these parameters, we choose to not tune these parameters.

\paragraph{Noise2Recon-SS.} The Noise2Recon-SS model was trained with training noise range $\noiserange=[0.2,0.5)$. Consistency loss weight was $\lambda=0.05$. Balanced sampling was not used as all training examples were unsupervised (i.e. $\mathcal{D}^{(s)}=\emptyset$). All examples in the batch were used in both the reconstruction and consistency pathways.

\subsection{Metrics and Losses}
For all experiments, we report results using three common image quality metrics -- the magnitude normalized root mean squared error (nRMSE, \cref{eq:nrmse}), structural similarity (SSIM) following the implementation from \cite{wang2004image}, and peak signal-to-noise ratio (pSNR, \cref{eq:psnr}). $\hat{x}$ is the complex-valued prediction and $x$ is the complex-valued reference.

\begin{equation}
\label{eq:nrmse}
    \text{nRMSE($\hat{x}$, $x$)} = \frac{||\;\;|\hat{x}| - |x|\;\;||_2}{||x||_2}
\end{equation}

\begin{equation}
\label{eq:psnr}
    \text{PSNR($\hat{x}$, $x$)} = 20\log_{10} \frac{\max |x|}{||\;\;|\hat{x}| - |x|\;\;||_2}
\end{equation}

We also include definitions for the image-space $\ell_1$ loss and k-space $\ell_1$ losses. $N_p$ refers to the number of pixels in the image. Note for the k-space $\ell_1$ loss, we do not scale by the number of pixels in the example.
\begin{equation}
\label{eq:l1-image-loss}
    \ell_{1, image}(\hat{x}, x) = \frac{||\hat{x} - x||_1}{N_p}
\end{equation}

\begin{equation}
\label{eq:l1-kspace-loss}
    \ell_{1, kspace}(\hat{x}, x) = ||A\hat{x} - Ax||_1
\end{equation}

\RV{
\subsection{Experimental Setup}
\label{app:experimental-setup}
\paragraph{Cross-dataset setup.} In this setup, models trained on the mridata 3D fast-spin-echo knee dataset were evaluated on the 4x-accelerated 2D fastMRI brain dataset. We evaluated different methods against examples in the 2D fastMRI brain validation multi-coil dataset that maximized the sources of distribution shift during evaluation. These shifts included:
\begin{itemize}[noitemsep]
    \item Anatomy: knee $\rightarrow$ brain
    \item Acceleration: 12x $\rightarrow$ 4x
    \item Scan Type: 3D $\rightarrow$ 2D
    \item Undersampling pattern: 2D Poisson Disc $\rightarrow$ 1D Random Undersampled
    \item Acquisition: 3D PD FSE $\rightarrow$ 2D T2, 2D FLAIR, 2D T1 Pre- and Post Contrast
    \item Field Strength: 3T $\rightarrow$ 1.5T
\end{itemize}
603 scans in the fastMRI multi-coil brain validation dataset contained the listed distribution shifts. Sensitivity maps for each volume were estimated using JSENSE (implemented in SigPy \cite{sigpy}) with a kernel-width of 8 and a 26$\times$26 center k-space auto-calibration region (equivalent to 8\% auto-calibration region) \cite{ying2007joint}. Fully-sampled data were retrospectively undersampled with a 1D random undersampling pattern with the same auto-calibration region. For testing, a unique, deterministic undersampling trajectory was generated for each testing volume using a fixed random seed for reproducibility.
}

\begin{table}[t!]
\begin{center}
\resizebox{\columnwidth}{!}{
\begin{tabular}{|c|l|c|c|c|c|c|}
\hline
$R$ \ Noise Level & 0 & 0.2 & 0.4 & 0.6 & 0.8 & 1.0 \\
\hline
12x & 0.07 & 0.15 & 0.3 & 0.6 & 0.9 & 1.2 \\
\hline

16x & 0.06 & 0.12 & 0.25 & 0.5 & 0.8 & 1.1 \\
\hline
\end{tabular}
}
\end{center}
\caption{Regularization parameter selection for compressed sensing at various noise levels with 12x and 16x acceleration ($R$).}
\label{tbl:cs-reg}
\end{table}

\begin{table}[t!]
    \centering
    \resizebox{\columnwidth}{!}{
    \begin{tabular}{llll}
    \toprule
    &           nRMSE &          SSIM &     pSNR (dB) \\
    \midrule
    $p$=0.1  &  0.139 (0.011) &  0.875 (0.012) &  39.3 (0.476) \\
    $p$=0.2* &  0.137 (0.011) &  0.889 (0.010) &  39.4 (0.486) \\
    $p$=0.3  &  0.136 (0.010) &  0.894 (0.009) &  39.4 (0.440) \\
    $p$=0.5  &  0.142 (0.009) &  0.861 (0.007) &  39.1 (0.342) \\
    \bottomrule
    \end{tabular}
    }
    \caption{The effect of augmentation probability $p$ on in-distribution performance ($\testnoise=0$) of supervised baselines trained with noise augmentations (Supervised+Aug). Highest performance is achieved at $p$=0.2,0.3. Asterisk (*) indicates the default augmentation probability used for baseline augmentation methods.}
    \label{tbl:supervised-aug-p}
\end{table}

\begin{table}[t!]
    \centering
    \resizebox{\columnwidth}{!}{
    \begin{tabular}{llcccc}
        \toprule
       &  & \multicolumn{2}{c}{$\testnoise=0.2$} & \multicolumn{2}{c}{$\testnoise=0.4$} \\
       \cmidrule(r){3-4} \cmidrule(r){5-6}
      $R$  & Method &     pSNR (dB) &           SSIM &     pSNR (dB) &           SSIM \\
    \midrule
12x & Noise2Recon (Ours) ($k$=1) &      36.7 &  \textbf{0.876} &      \textbf{35.7} &  \textbf{0.852} \\
12x & Compressed Sensing ($k$=1) &      35.5 &  0.845 &      33.2 &  0.743 \\
 12x  & Supervised + Aug ($k$=1) &      35.7 &  0.765 &      34.7 &  0.748 \\
12x   & Supervised + Aug (FT) ($k$=1) &      36.1 &  0.765 &      35.0 &  0.746 \\
 12x  & Supervised + Aug ($k$=14) &      \textbf{36.9} &  0.851 &      35.7 &  0.803 \\
 12x  & Supervised + Aug (FT) ($k$=14) &      37.0 &  0.865 &      35.6 &  0.804 \\
\midrule
16x  & Noise2Recon (Ours) ($k$=1) &      36.6 &  \textbf{0.862} &      35.4 &  \textbf{0.838} \\
16x & Compressed Sensing ($k$=1) &      35.1 &  0.827 &      32.7 &  0.707 \\
 16x  & Supervised + Aug ($k$=1) &      35.7 &  0.784 &      34.6 &  0.746 \\
 16x  & Supervised + Aug (FT) ($k$=1) &      35.8 &  0.755 &      34.7 &  0.734 \\
  16x & Supervised + Aug ($k$=14) &     \textbf{36.7} &  0.856 &      \textbf{35.5} &  0.798 \\
 16x  & Supervised + Aug (FT) ($k$=14) &      36.7 &  0.847 &      35.5 &  0.802 \\
    \bottomrule
    \end{tabular}
    }
    \caption{Reconstruction performance in low-SNR settings ($\testnoise>0$). With only one supervised training example ($k$=1), Noise2Recon outperforms supervised DL baselines by over 1dB pSNR and 10\% SSIM. It also matches performance of augmentation-based supervised networks trained with 14 supervised scans ($k$=14). Thus, Noise2Recon may be a data-efficient and OOD-robust alternative to existing CS and DL methods. Values are equivalent to data in Fig. \ref{fig:noisy-line-graph}.}
    \label{tbl:noisy-performance}
\end{table}

\begin{figure}[t!]
  \centering
  \begin{center}
      \includegraphics[width=\linewidth]{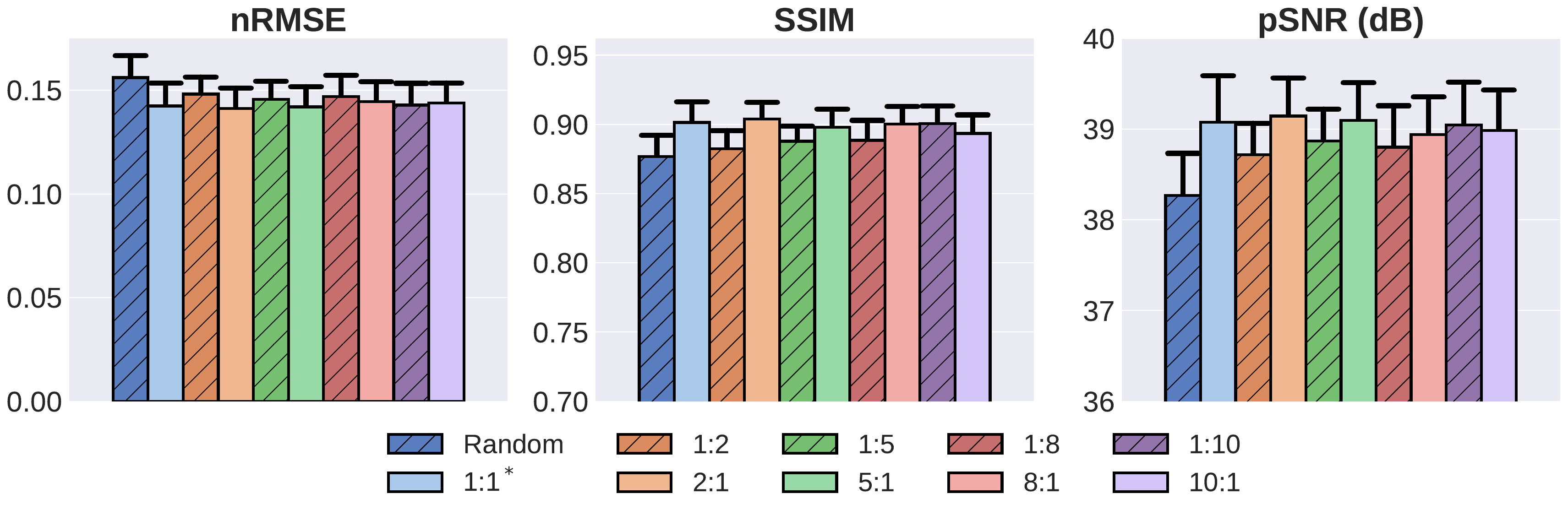}
  \end{center}
  \caption{Balanced sampling of supervised to unsupervised scans compared to random sampling. Asterisk (*) indicates the default sampling configuration for Noise2Recon experiments. Balanced sampling, regardless of the ratio of supervised to unsupervised \RV{($T_s:T_u$)} examples, increases average performance over standard random sampling.}
  \label{fig:abl-balanced-sampling}
\end{figure}

\section{Additional Experimental Results}
\label{app:additional-exps}
This section provides details regarding additional experiments. All models were trained with the following configurations unless otherwise noted. Noise2Recon models were trained with 1 supervised training subject and 13 unsupervised training subjects with 1:1 balanced sampling between supervised and unsupervised scans. Consistency loss weight was $\lambda=0.1$. During training, the noise level was sampled at random from the specified range $\noiserange$. Supervised+Aug models were trained with 14 supervised training subjects with 20\% probability (p=0.2) of applying augmentations.

\begin{figure}[t!]
  \centering
  \begin{center}
      \includegraphics[width=\linewidth]{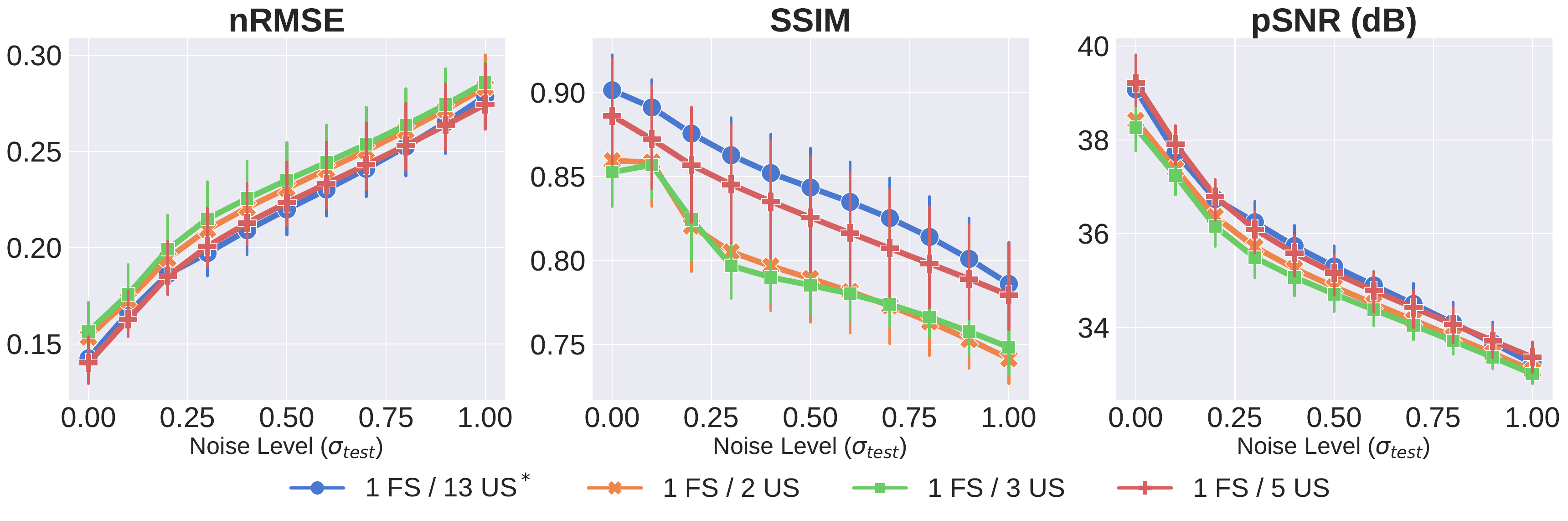}
  \end{center}
  \caption{Noise2Recon performance at multiple noise levels ($\testnoise$) with increasing number of undersampled (US) examples. Notation \textit{A/B} denotes \textit{A} fully-sampled (FS, i.e. supervised) scans and \textit{B} undersampled (i.e. unsupervised) scans for training. Asterisk (*) indicates the default FS / US ratio configuration for experiments. Increasing the number of US examples improved performance at all noise levels, which may suggest that Noise2Recon is stable in cases of large imbalances in the number of FS and US examples.}
  \label{fig:increasing-num-unsupervised}
\end{figure}

\begin{figure}[t!]
  \centering
  \begin{center}
      \includegraphics[width=\linewidth]{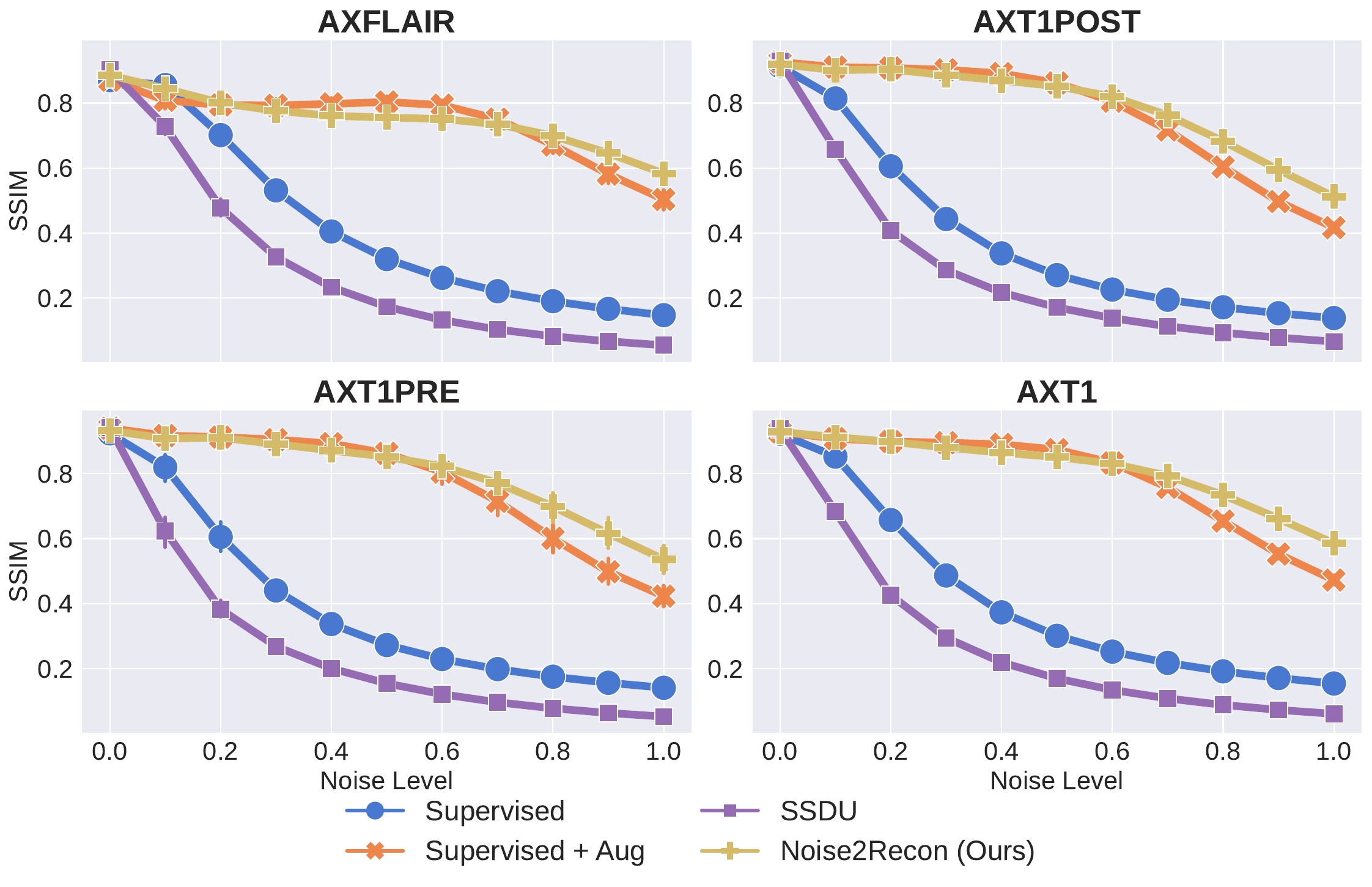}
  \end{center}
  \caption{Cross-dataset generalizability (mridata knee $\rightarrow$ fastMRI brain) with unrolled networks. Noise2Recon performs comparably to SSDU among in-distribution (high-SNR) data and comparably to augmentation methods in OOD (low-SNR) regimes.}
  \label{fig:fastmri-unrolled-noise}
\end{figure}

\subsection{Scaling with Increasing Unsupervised Data}
In practice, the undersampled-only (unlabeled) scans are more prevalent and collected more frequently than fully-sampled (labeled) scans. In this ablation, we explore the impact of increasing the number of undersampled examples during training. Models were trained with 1 fully-sampled scan and 2, 3, 5, or 13 undersampled scans. Noise2Recon performance improved as the number of undersampled scans used for training increased (Fig. \ref{fig:increasing-num-unsupervised}). The increased performance with larger undersampled datasets may indicate that Noise2Recon is robust to size imbalances in supervised and unsupervised datasets. As the framework relies on pseudo-label generation, this observation may suggest that the quality of pseudo-labels improves with more undersampled scans. 


\subsection{Supervised Augmentation Probability \textit{p}}
\label{app:baseaug-noise-prob}
In this ablation, we measure the effect of augmentation probability on supervised training with noise augmentations (Supervised+Aug). Supervised baselines were trained with noise augmentations applied with probabilities of $p$=0.1, 0.2, 0.3, and 0.5. Highest performance was observed at configurations $p$=0.2 and $p$=0.3 (Table \ref{tbl:supervised-aug-p}). Augmentation probability $p$=0.2 was selected as the default configuration for training all supervised methods with noise augmentations.

\subsection{Sample Reconstructions Under Real Noise}
\begin{figure}[t!]
  \centering
  \begin{center}
      \includegraphics[width=\linewidth]{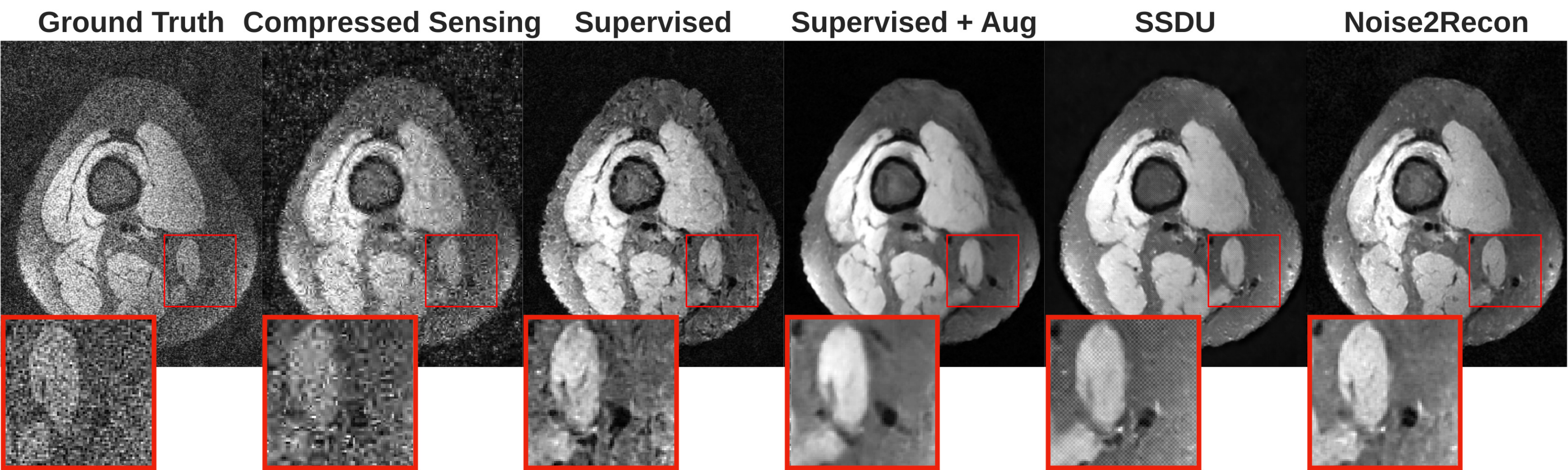}
  \end{center}
  \caption{Sample reconstruction from an edge slice for different methods at acceleration $R$=12. When presented with a noisy, undersampled image at inference time, Noise2Recon jointly performs denoising and reconstruction to recover anatomies that were acquired with a low-SNR during acquisition.}
  \label{fig:edge-slice}
\end{figure}

In 3D scans, SNR profile can change based on the spatial encoding of the slice. To assess the performance of Noise2Recon in reconstruction from a low-SNR image, we visualized an edge slice from the test set where the inherent noise observed during acquisition for the ground-truth was higher compared to middle slices. We observed that Noise2Recon can produce robust reconstructions regardless of spatially-localized SNR differences (Fig. \ref{fig:edge-slice}).

\section{Sample Zero-Filled Reconstructions of Noisy Images}
\label{app:train-noise-range}
\begin{figure}[t!]
  \centering
  \begin{center}
      \includegraphics[width=\linewidth]{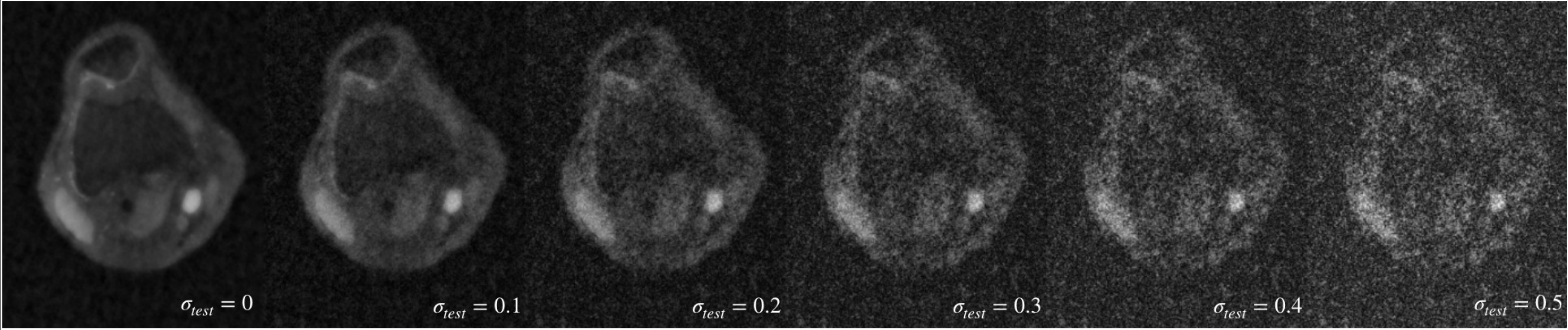}
  \end{center}
  \caption{Zero-filled, SENSE-reconstructed images are shown at acceleration rate $R = 12$ under various noise levels $\testnoise \in \{0, 0.1, 0.2, 0.3, 0.4, 0.5\}$. }
  \label{fig:zf-noise}
\end{figure}

In our experiments, $\testnoise$ was varied from 0, 0.1, $\dots$, 1.0. On a representative test knee slice, we demonstrate the impact of noise on zero-filled, SENSE-reconstructed images at acceleration rate $R = 12$ for noise levels $\testnoise \in \{0, 0.1, 0.2, 0.3, 0.4, 0.5\}$ in Fig. \ref{fig:zf-noise}.

\section{Additional Discussion}
\paragraph{Compressed sensing and denoising} In CS, recovering images from undersampled measurements is made possible by introducing incoherence through random undersampling, where resulting aliasing artifacts resemble additive Gaussian noise \cite{cs-mri}. As a result, image recovery with CS can be viewed as a denoising problem. Noise2Recon aims to utilize the similarity between the reconstruction and denoising tasks by jointly optimizing a reconstruction and denoising objective. We observe that optimizing similar tasks jointly helps, where performance improves both in the reconstruction task (Fig. \ref{fig:acc-one-to-many}), and in the denoising task (Fig. \ref{fig:noisy-line-graph}). Our observations are in agreement with multi-task learning theory, which suggests that given similar tasks, optimizing a multi-task objective leads to positive transfer \cite{wu2020understanding}. Positive transfer refers to improving performance on a task by training a joint objective of multiple tasks, compared to training a task individually. 


\paragraph{Task-based regularization} Our joint reconstruction and denoising paradigm is reminiscent of model regularization techniques. Traditionally, these methods are designed to reduce the variance of the model by convex constraints such that the model parameters are sparse or low-magnitude \cite{larsen1994generalization,van2017l2}. The addition of the denoising objective may not regularize the network at the parameter-level, but rather at the more semantic task-level. With the consistency objective, the regularizer is explicitly data-driven, which may help learn non-convex regularization processes that are optimal for the collected data.

\end{document}